\documentclass[pre,twocolumn,aps,showpacs]{revtex4}
\usepackage{graphicx}
\usepackage[activeacute,english]{babel}

\begin{document}

\title{Lower bounds on dissipation upon coarse graining}

\author{A. Gomez-Marin$^1$, J. M. R. Parrondo$^2$ and C. Van den Broeck$^3$}

\address{$^1$ Facultat de F\'isica, Universitat de Barcelona,
Diagonal 647, 08028 Barcelona, Spain \\ $^2$ Departamento de F{\'i}sica
At\'omica, Molecular y Nuclear and {\em GISC}, Universidad
Complutense de Madrid, 28040 Madrid, Spain \\
$^3$ Hasselt University, B-3590 Diepenbeek, Belgium}


\begin{abstract}
By different coarse-graining procedures we derive lower bounds on
the total  mean work dissipated in Brownian systems driven
out of equilibrium. With several analytically solvable examples we
illustrate how, when and where the information on the dissipation is
captured.
\end{abstract}


\pacs{05.70.Ln, 05.40.-a}

\maketitle

\section{Introduction}

Equilibrium statistical physics provides the microscopic foundation
of thermodynamics, built around the concept of {\it entropy} as the
logarithm of the phase volume. The theory has been extended to the
regime of linear irreversible thermodynamics by identifying the {\it
entropy production} in the regime of linear response \cite{onsager,onsager2,prigogine,prigogine2}. There exists to date no general theory
covering far from equilibrium situations.
However, recent
results known as fluctuation
\cite{fluctuation,fluctuation2,fluctuation3,fluctuation4,fluctuation5}
or work \cite{work,worka,workb,workc,work2,work2a,work3,work4,work5} theorems
point to the
existence of exact equalities valid independent of the distance from
equilibrium. These equalities involve fluctuations in work or
entropy production.
For the average of these quantities, they reduce
to inequalities, in agreement with the second law of thermodynamics.
For example, the Jarzynski equality states that $\langle \exp(-\beta
W) \rangle = \exp (-\beta \Delta F)$, where $W$ is the work needed
to bring a system, in contact with a heat bath at temperature $T$
($\beta^{-1}\equiv k_B T$) from one initial equilibrium state
to a final one, and $ \Delta F$ is the difference in free energy of
these states (see  \cite{work2} for a more precise discussion). Then by
the application of Jensen's inequality one finds $\langle W \rangle
\ge \Delta F$. While the work and fluctuation theorems are certainly intriguing
results of specific interest for the study of small systems, they
provide no extra information on the average value of work and
entropy production.

Parallel to these developments, the direct calculation of work,
entropy and dissipation has proceeded following various channels of
research. Explicit results for path dependent dissipation or entropy
production have been obtained, mostly in the context of Markovian
stochastic models \cite{crooks,maes,seifert2005,gaspard2004}.
The  average entropy production goes back to earlier work
\cite{schnakenberg,luo}. Recently the microscopically exact value of
the average dissipated work has been obtained in a set-up similar to
that of the work theorem \cite{diss}.
The connection and consistency  between this exact Hamiltonian
result and the ones derived in the context of stochastic models has been clarified in \cite{bounds}.

The main issue of this paper is to discuss how this type of formulas
can be applied when only limited information on the system is
available. In other words how much of the irreversible behavior is
revealed when only partial measurements are performed. Complimentary
to this issue is the identification of the variables in which the
traces of the dissipation reside. One could imagine that such
knowledge would allow to develop mechanisms to either increase or
decrease the dissipation. The limitations in observation can be of
different nature. It could be that the monitoring in time is not
exhaustive, but one only performs a finite number of punctual
measurements. The measurement of the variables could also be crude
or incomplete. Or both limitations could be present. We will show that
this limited information provides a lower bound for the dissipation.
Of particular interest is to know whether this bound is close to the
full dissipation or just reproduces the general bound that is
contained in the second law. As such, the analysis will reveal when
and where the information on the dissipative process is located. For
our illustration, we will focus on systems described by stochastic
dynamics. The latter provide an accurate description of  mesoscopic
phenomena in physics, chemistry and biology, and have been used
extensively especially in the context of Brownian entities appearing
in nano and bio-technology. The additional advantage is that
calculations can be carried out in full analytic detail.

The lay-out of the paper is as follows. We start with a discussion
of  the basic expression  for dissipation in terms of relative
entropy and its relation to the arrow of time.  Parallel to these developments, this result appears in the framework of both  microscopic analysis \cite{diss,bounds,jar06,jar07}
and  stochastic thermodynamics
\cite{crooks,schnakenberg,gaspard2004,luo,maes,seifert2005,bly08} for transient
and steady nonequilibrium states. We explain how this expression can
be used to bound dissipation from below. We next investigate in a
number of experimentally relevant examples, how these bounds  apply
in the case of coarse-graining applied to the measurement in time,
to the choice of variables, or to both. The illustrations include overdamped and underdamped Brownian particles in moving and quenched potentials.

\section{Relative entropy, coarse-graining and lower bounds}

We introduce the quantity $\Delta s$ which measures the
irreversibility in a path realization $z(t)$ for an arbitrary
stochastic process, in extension of its definition for Markov
processes \cite{crooks,seifert2005}, continuous time random walks
 or dynamical  systems
\cite{gaspard2004,gaspard2007}:
\begin{equation} \label{RR}
\Delta s \equiv k_B \ln  \frac{
\mathcal{P}[z(t)]}{\tilde{\mathcal{P}}[\tilde{z}(t)]}.
\end{equation}
Here  $\mathcal{P}$ is the probability of  observing the so-called
forward path ${z}(t)$. The tildes refer to the time-reversed analogue.
$\tilde{z}(t)=z(t_f-t)$ is the time-reversed trajectory,
in which the sign of the momenta are reversed if such variables appear in the
description.
$t_f$ is the total duration of the nonequilibrium experiment.
$\tilde{\mathcal{P}}$ is then the probability for such
a trajectory in an experiment employing the time-reversed schedule
of the perturbation.
The study of the above trajectory-dependent
quantity itself is of considerable interest \cite{seifert2005,gaspard2007}.

In this paper, we are only interested in its average  (the notation
suggests continuous variables, but the results are
trivially reproduced for discrete dynamics):
\begin{equation}  \label{Raver}
\langle \Delta s \rangle
= k_B\int \mathcal{D}z(t)
\mathcal{P}[z(t)] \ln
\frac{ \mathcal{P}[z(t)] }
{ \tilde{\mathcal{P}}[\tilde{z}(t)] } \equiv  k_B D(\mathcal{P}||\tilde{\mathcal{P}}).
\end{equation}
This average quantity is expressed in terms of the {\it relative entropy} $D(\mathcal{P}||\tilde{\mathcal{P}})$ (also called Kullback-Leibler distance) between the distributions $\mathcal{P}$ and $\tilde{\mathcal{P}}$.
The relative entropy has a number of  extremely powerfull and useful properties. In particular, it is a positive quantity whose value decreases upon any type of {\it coarse-graining} \cite{cover}.

More precisely, if the statistical information on the detailed path trajectory $z(t)$
 (which could be generically decomposed in two subsets as $z\equiv \{x,y \}$)
is not available, one considers the reduced trajectory
$z_{\rm{cg}}\equiv x$, where the subscript refers to
coarse-graining. Then
\begin{eqnarray}
& D(\mathcal{P}||\tilde{\mathcal{P}})\equiv
D(\mathcal{P}(z)||\tilde{\mathcal{P}}(\tilde{z}))   =
   \int
dx\,dy\, \mathcal{P}(x,y)\ln\frac{\mathcal{P}(x,y)}{
\tilde{\mathcal{P}}(\tilde{x},\tilde{y})
} &
\nonumber \\
 & =
 D(\mathcal{P}(x)||\tilde{\mathcal{P}}(\tilde{x}) )   +   \int\!  dx  \,\mathcal{P}(x) \!\int\!  dy\,
\mathcal{P}(y|x)\ln\frac{\mathcal{P}(y|x)}{\tilde{\mathcal{P}}(\tilde{y}|\tilde{x})}
&
\nonumber \\
& \ge  D( \mathcal{P}(x)|| \tilde{\mathcal{P}}(\tilde{x}) )   \equiv
 D( \mathcal{P}_{\rm{cg}}|| \tilde{\mathcal{P}}_{\rm{cg}} ),&
 \label{chainrule}
\end{eqnarray}
where $\mathcal{P}_{\rm{cg}}$ is the corresponding coarse-grained probability.
Finally,
in combination with (\ref{Raver}), we obtain:
\begin{eqnarray} \label{Rcg1}
\langle \Delta s \rangle
&= &
k_B D(\mathcal{P}||\tilde{\mathcal{P}}) \ge
k_B D(\mathcal{P}_{\rm{cg}}||\tilde{\mathcal{P}}_{\rm{cg}})
\geq 0.
\end{eqnarray}
Notice that so far no assumption has been made on the stochastic
dynamics. The above expressions  are valid for example for deterministic systems (with
distributed initial conditions) and for non-Markovian processes.
The
quantity $\Delta s $ is commonly regarded in many scenarios as the
trajectory-dependent total entropy production. Thus, the average of
$\Delta s$ over the ensemble of trajectories corresponds to the
total  thermodynamic entropy production $\Delta S$. From the
relative entropy properties, equation (\ref{Rcg1}) conveys more
information than the second law of thermodynamics itself, namely,
that the total entropy production is always greater than zero and,
furthermore, that the better our description, the more precisely one
can estimate the actual total dissipation approaching from below.
Equation (\ref{Rcg1}) puts together two interesting concepts: the
intrinsic nature of irreversibility and the subjective partial
information due to incompleteness of measurements.

In the case in which the system is initially prepared in equilibrium
and then a  transient nonequilibrium excursion takes place, the
mean total entropy production comes in the form of mean work dissipated,
$\langle W_{\rm{diss}} \rangle$, which combined with (\ref{Rcg1})
implies
\begin{eqnarray} \label{WdissCG}
\langle W_{\rm{diss}}\rangle \equiv \langle W\rangle -\Delta F  \ge
k_B T D(\mathcal{P}_{\rm{cg}}||\tilde{\mathcal{P}}_{\rm{cg}})
\geq 0.
\end{eqnarray}
This equation has been proved exactly for Hamiltonian dynamics in
\cite{diss}  and formally extended to mesoscopic descriptions in
\cite{bounds}.
The main goal of this work is to apply the above formula
to different Brownian systems driven out of equilibrium in which the
effects of coarse-graining can be illustrated. We present three
different analytically solvable examples, all initially
prepared in equilibrium.
First we consider an
overdamped Brownian particle in a moving  trap
\cite{ritort,gaspard2007} whose trajectory is coarse-grained in
time.
Second, we introduce an underdamped Brownian particle in a
suddenly changing stiffening trap. We measure its relative entropy
at one point in time after the quench and then we integrate out the
position and momentum variables. Third, by means of two linearly
coupled underdamped Brownian particles, we study the flow of
information on dissipation amongst several degrees of freedom as
a function of time. Lower bounds for the total mean work dissipated
are derived in these three first examples.

\section{Study cases}

\subsection{Overdamped Brownian particle in a constant-speed moving trap}\label{sec:over}

In this section we present a solvable example, which is moreover of
experimental relevance,  namely, an overdamped Brownian particle
subject to  a moving time-dependent harmonic potential:
\begin{equation}\label{mhp}
V(x,t)=\frac{k}{2}(x-ut)^2,
\end{equation}
where $k$ is the stiffness of the trap and $u$ is the  constant
velocity at which the trap is moved. The time evolution of the
position variable  $x$ of  the overdamped particle obeys the
following Langevin equation
\begin{equation}\label{lang}
\dot{x}=-\partial_x V(x,t)+\xi(t).
\end{equation}
$\xi(t)$ is a Gaussian white noise, with $\langle
\xi(t)\xi(t')\rangle=2T \delta(t-t')$. For simplicity of notation,
we have absorbed the friction coefficient in the time unit and the
Boltzmann constant $k_B$ in the definition of temperature.

Before proceeding to the relation between dissipation  and relative
entropy, we review the salient features of the energy balance. Our
starting point is conservation of total energy, or first law, at the
level of a single stochastic trajectory \cite{seki}, during an
experiment from initial time $0$ to final time $t_f$. Since the
particle is instantaneously thermalized at the constant temperature
$T$ of the heat bath, its change in energy is equal to its change in
potential energy $\Delta V=V(x(t_f),t_f)-V(x(0),0)$. The latter must
be equal to the amount of work $W$ exerted by the external force
(sometimes called the injected work) minus the heat $Q$ delivered to
the heat bath (also referred to as dissipated heat to the
environment):
\begin{equation} \label{1stLaw}
\Delta V=\int_{0}^{t_f} \frac{dV}{dt}dt= \int_{0}^{t_f}
\frac{\partial V}{\partial t}dt +\int_{0}^{t_f} \frac{\partial
V}{\partial x} \dot{x}dt=W-Q.
\end{equation}
From such energy balance the fluctuating heat and work can be
identified \cite{sekimoto}: the rate of heat dissipated to the heat
bath is given by $\dot{Q}=-\partial_x {V} \dot{x}$,  while the work
done per unit time in moving the external potential is
$\dot{W}=\partial_t{V}$. These quantities depend on the actual
realization of the stochastic trajectory $x(t)$. Thus heat and work
are random variables. The fact that injected work and dissipated
heat differ by the energy stored in the particle has important
consequences for their large deviation properties for asymptotically
large times when the latter energy is unbounded. The fluctuation
theorem has therefore to be carefully reconsidered
~\cite{farago,vanzon,blickle,hfbt,visco,maes2006,joubaud}.

We are concerned here with the average work,  in which case large
deviation issues are irrelevant. Using the explicit expression of
the potential (\ref{mhp}), one finds
\begin{eqnarray}
\langle W \rangle &=& \left\langle \int_{0}^{t_f}\frac{\partial
V(x,t)}{\partial t}  dt  \right\rangle = \left\langle \int_{0}^{t_f}
dt k(x-ut)(-u)  \right\rangle \nonumber \\ &=& u \int_{0}^{t_f} dt
\langle \dot{x}(t)-\xi(t) \rangle =u [\langle x(t_f)\rangle -\langle
x(0) \rangle].
\end{eqnarray}
On the other hand, the average of equation (\ref{lang}) yields the
following exact closed equation for the average position
\begin{equation}\label{langeq}
{\langle \dot{x} \rangle} = -\langle \partial_x V \rangle=-k(\langle x \rangle -ut),
\end{equation}
whose solution for an arbitrary initial condition at
$t_0$  will be useful later on:
\begin{equation}\label{mean}
{\langle x(t) \rangle} =e^{-k(t-t_0)}\langle x(t_0)
\rangle+\frac{u}{k} [kt -1-e^{-k(t-t_0)}(kt_0-1)].
\end{equation}
If the system is prepared initially in equilibrium, from (\ref{lang}) it is clear that
$\langle x(0) \rangle=0$. The translation of the harmonic
potential does not change the free energy of the system, $\Delta
F=0$.  Then the dissipative work equals in average the external
work:
\begin{equation} \label{Wdiss}
\langle W_{\rm{diss}} \rangle \equiv \langle W \rangle- \Delta F=
\frac{u^2}{k} (kt_f +e^{-kt_f}-1).
\end{equation}

In the sequel, we will illustrate how Eq. (\ref{WdissCG}) approaches
to the exact dissipative work $(\ref{Wdiss})$ from below as we
include in the calculation of the relative entropy more information
on the paths. We will consider, as would occur in an experimental or
numerical realization of our example, that the position $x(t)$ is
measured only at a finite instants of time. This information loss
about the path can be viewed as a coarse-graining in time. If the
relative entropy is calculated with this partial information, Eq.
(\ref{WdissCG}) will give us only a rigorous lower bound. The
calculation which we are about to perform will tell us how fast this
bound converges to the exact value.

\begin{figure}
\begin{center}
\includegraphics[angle=0, width=8cm]{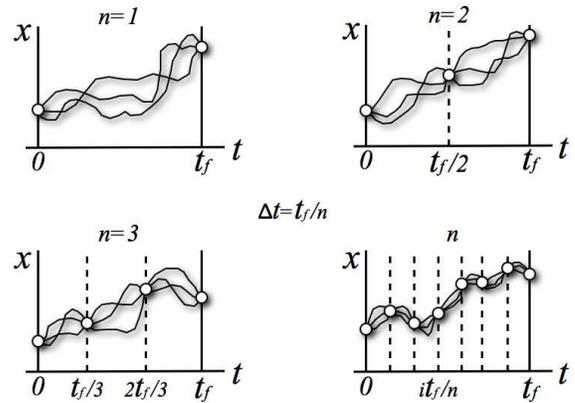}
\caption{Sketch of the $n$-slicing procedure in which the full
trajectory of the particle is not measured but only its position
after time intervals $\Delta t =t_f/n$, where $t_f$ is the total
duration of the experiment. } \label{scheme}
\end{center}
\end{figure}

For simplicity we will consider that the coarse graining is into $n$
equal divisions $\Delta t\equiv t_f/n$  of the total time duration
$t_f$. Therefore, in this $n$-slicing procedure, the full trajectory
of the particle  is not measured but only its position after time
intervals of duration $\Delta t$. See figure \ref{scheme}. The
probability for a discretized path can be easily evaluated since the
process is Markovian and Gaussian. Let us denote by $p(x_{i+1}|x_i)$
the conditional probability for jumping from a point $x_i$ at time
$t_i$ to a point $x_{i+1}$ at time $t_i+\Delta t$, and let
$p^{\rm{eq}}_{0}$ be the initial equilibrium distribution. The
probability $\mathcal{P}_{\rm{cg}}$ of the $n$-sliced discretized
path
$\vec{x}\equiv[x_0,\;x_{1},\;...,\;x_{i},\;...,\;x_{n-1},\;x_{f}]$
 is given by
\begin{eqnarray}\label{markovian}
\mathcal{P}_{\rm{cg}} & \equiv & \mathcal{P}_{\rm{cg}}([x_0,\;x_{1},\;...,\;x_{i},\;...,\;x_{n-1},\;x_{f}]) \nonumber \\
&= &
p^{\rm{eq}}_{0}(x_0) \prod_{i=0}^{n-1} p(x_{i+1}|x_i).
\end{eqnarray}
An analogous expression is valid for the backward path and
probability, with superscript ``tilde'' again referring to time
reversed excursion (trajectory and process).
The central quantity we wish to evaluate is the
following coarse-grained relative entropy $I_n$;
\begin{eqnarray} \label{sumIn}
I_n & \equiv & T D(\mathcal{P}_{\rm{cg}}(\vec{x})||\tilde{\mathcal{P}}_{\rm{cg}}(\tilde{\vec{x}})) \nonumber \\
& = &
T\left\langle  \ln \frac{p^{\rm{eq}}
_0(x_0)}{p^{\rm{eq}}
_f(x_f)} \right\rangle +
T\sum^{n-1}_{i=0} \left\langle \ln \frac{p(x_{i+1}|x_i)}{\tilde{p}(x_i|x_{i+1})} \right\rangle.
\end{eqnarray}
Note that we have multiplied by $T$ (having absorbed $k_B$ in its
units)  since we want to compare the above expression with the
dissipated work. The brackets $\langle ... \rangle$ refer to the
average performed with the forward distribution, which
weights every trajectory's contribution.

The next step is to find the general expression for $p(x_{i+1}|x_i)$
and $\tilde{p}(x_i|x_{i+1})$.
Since the Langevin equation that
describes the    dynamics is linear,
the conditional probabilities are Gaussian distributions:
\begin{equation} \label{pF}
p(x_{i+1}| x_{i})= \frac{1}{\sqrt{2\pi \sigma^2}} \exp \left[-\frac{
(x_{i+1}- \langle x_{i+1}\rangle_{x_i} )^2
  }{2\sigma^2}\right]
\end{equation}
and
\begin{equation} \label{pB}
\tilde{p}(x_{i} |x_{i+1})= \frac{1}{\sqrt{2\pi \sigma^2}}  \exp \left[-\frac{
(x_{i} -\langle \tilde{x}_{i}\rangle_{x_{i+1}})^2
  }{2\sigma^2}\right].
\end{equation}
From equation (\ref{mean}) (applied for final and initial times
$t_{i+1}$ and $t_i$, respectively, and with the appropriate initial
condition) the conditional averages are found to be
 \begin{eqnarray}
\langle x_{i+1}\rangle_{x_i} & = & x_i e^{-k\Delta t}+\omega+ \eta \;t _i,
\\
\langle \tilde{x}_{i}\rangle_{x_{i+1}} & = & x_{i+1} e^{-k\Delta t}-\omega+\eta \;t _{i+1}
\end{eqnarray}
where
\begin{equation}
\omega \equiv \frac{u}{k}(e^{-k \Delta t}+k\Delta t -1), \; \; \eta \equiv u(1-e^{-k \Delta t}).
\end{equation}
Similarly, one can multiply the Langevin equation by the position
$x$ and then take averages. This leads to the following equation for
the variance $\sigma^2 \equiv \langle x^2\rangle-\langle x \rangle^2$:
\begin{equation}
\frac{1}{2}\frac{d}{dt}\sigma^2=-k\sigma^2 +T,
\end{equation}
which yields (conditional variances starting at zero value)
\begin{equation}
\sigma^2=\frac{T}{k}(1-e^{-2 k \Delta t}),
\end{equation}
for both (forward and backward) cases.

In order to obtain $I_n$, we insert the above conditional
probability distributions in Eq.~(\ref{sumIn}). The final result can
most revealingly be written in terms of the duration of the
experiment $t_f$ and the final position of the minimum of the trap
$z_0\equiv u t_f$. After some cumbersome calculations, one finally
gets
\begin{equation} \label{In!}
I_n=\frac{z^2_0}{k t^2_f} \left[ e^{-kt_f}-1+2n \tanh \left( \frac{k t_f}{2 n} \right)
\right].\end{equation}

\begin{figure} [t]
\begin{center}
  \includegraphics[angle=270, width=7.7cm]{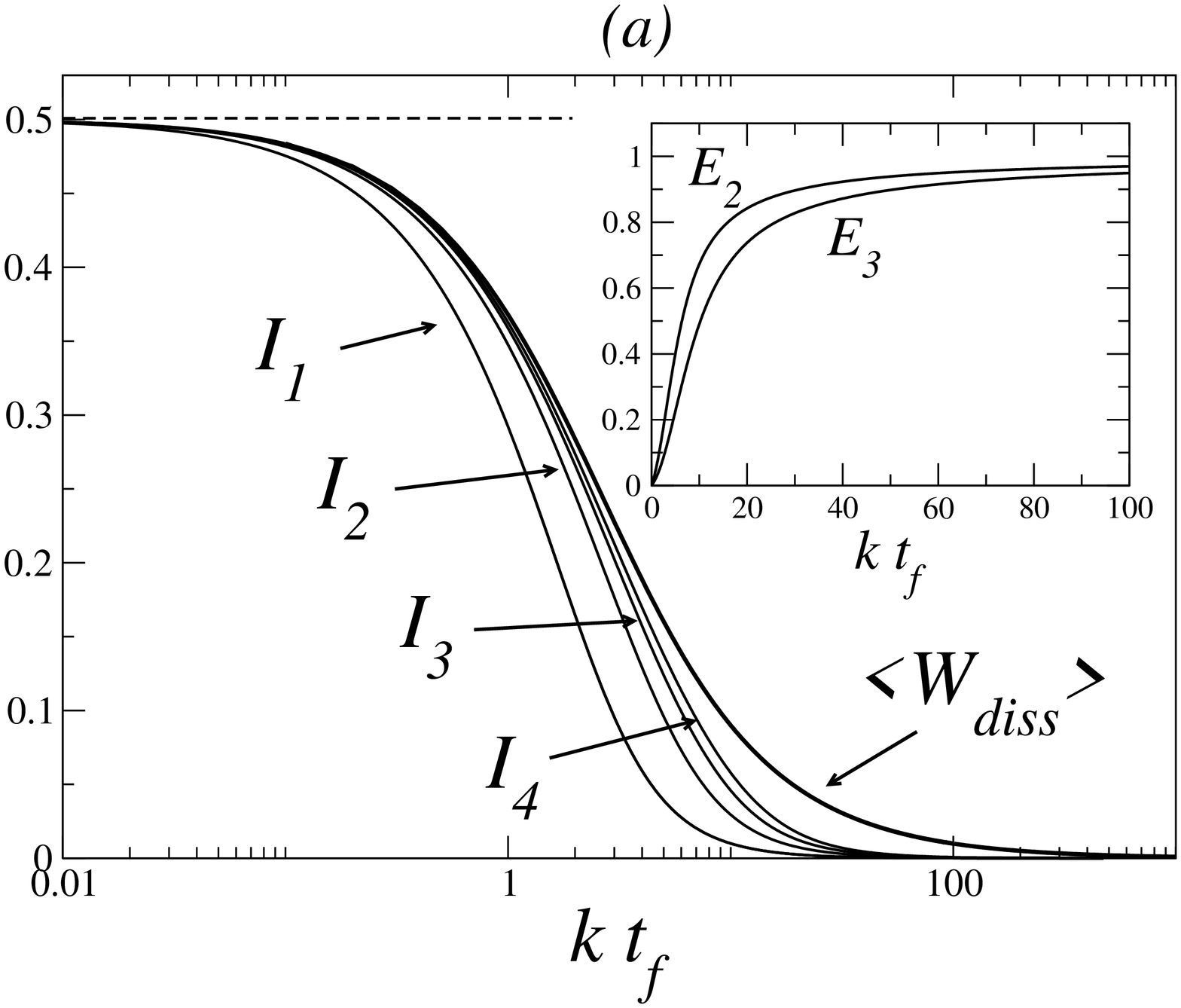}
   \includegraphics[angle=270, width=7.7cm]{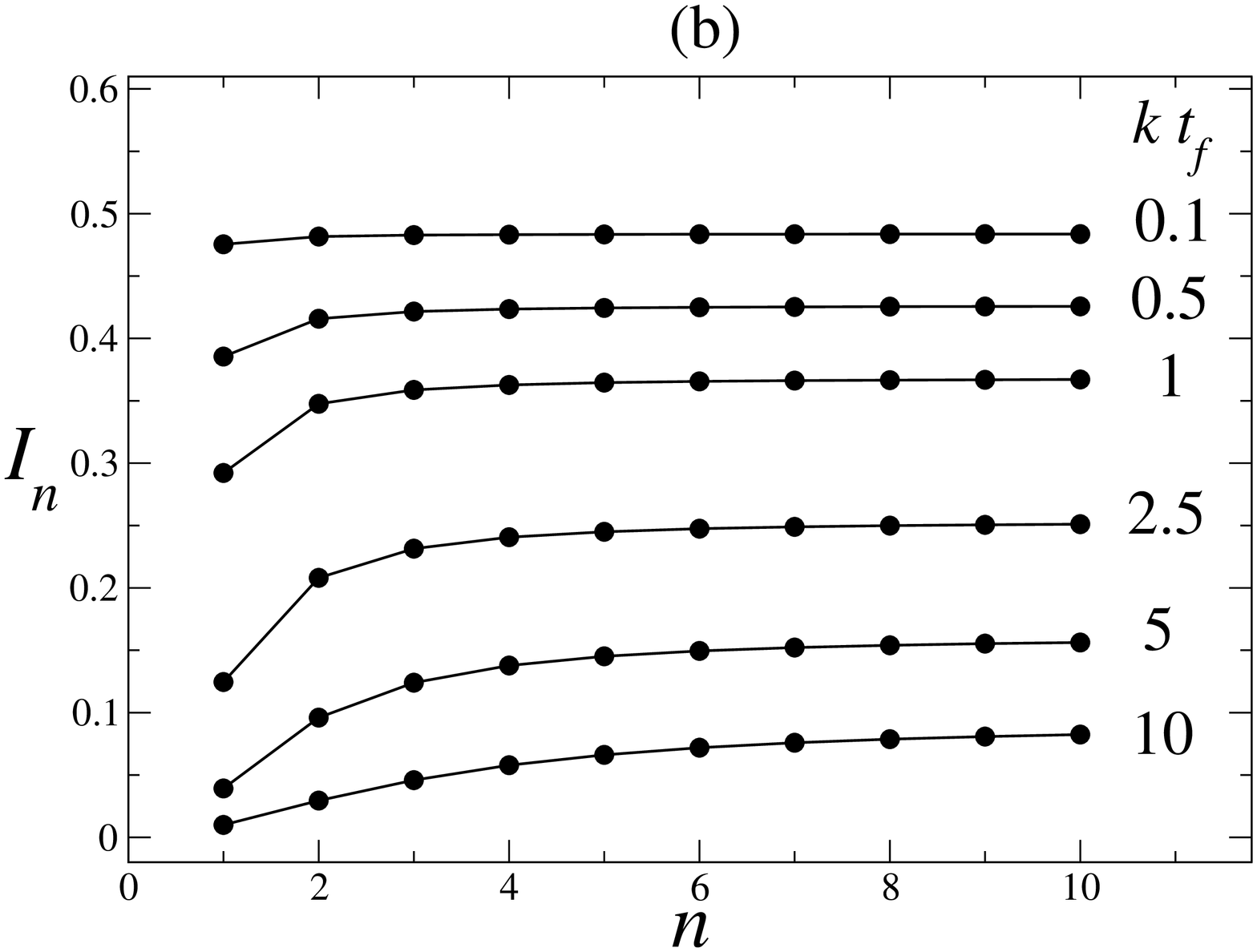}
  \caption{(a) Plot of $I_n$ (for $n=1,2,3,4$ and
  $I_\infty=\langle W_{\rm{diss}}\rangle$), as a function of the ratio of
  characteristic times $k t_f $. We have scaled out
  the prefactor $k z^2_0$. Note that  $I_n$ is always a lower bound to
  $\langle W_{\rm{diss}}\rangle$ and converges to the  irreversible instantaneous
  quench value (dashed line) and to the quasi-static limit (zero value) for  $k t_f \rightarrow 0$ and
  $k t_f \rightarrow \infty$, respectively. Inset:  the relative error
  $E_n\equiv (\langle W_{\rm{diss}}\rangle-I_n)/\langle W_{\rm{diss}}\rangle$ increases as a function of
  $k t_f $.
  (b) Plot of $I_n$  (for different values of $kt_f$) as a function of the number
  of time divisions $n$ of the trajectory.}
\label{boundsIn}
\end{center}
\end{figure}

First note that in the limit $n\to \infty$ one finds (cf.
Eq.~({\ref{Wdiss}))
\begin{equation} \label{I8}
I_{\infty}= \frac{ z^2_0}{kt_f^2}(kt_f +e^{-kt_f}-1)=\langle
W_{\rm{diss}}\rangle.
\end{equation}
Hence the exact dissipation is, as anticipated, recovered in the
limit of the continuous path description. We now turn to the
following question. How is the convergence of  $I_n$ to $\langle
W_{\rm{diss}}\rangle$? First,  one can verify that, for any value
of the system's parameters, $I_n$ is always a lower bound for the
total dissipation (cf. Eq. (\ref{WdissCG})):
\begin{equation}
\langle W_{\rm{diss}} \rangle \geq  I_n \geq 0.
\end{equation}
Next, as is apparent from the explicit result (\ref{In!}) and
scaling $k z_0^2$, the convergence of $I_n$ to $\langle W_{\rm
diss}\rangle$ depends only on the ratio of the time of the
experiment $t_f$ over the relaxation time $1/k$ in the harmonic
potential. In figure \ref{boundsIn}.(a) we plot $I_1$ up to $I_4$,
as a function of $k t_f$. The convergence is surprisingly good. For
example, for $k t_f=1$, the error in $I_2$ (single intermediate
measurement point, plus the initial and the final points, which are
always measured) is only a few percent.

In the  limit $u \rightarrow 0$ (or $k t_f \rightarrow \infty$),
that is, for a very slow translation of the potential,  one recovers
the quasi-static result of zero dissipated work. Note however that
the relative rate of convergence becomes quite bad in this limit
(cf. inset in figure \ref{boundsIn}.(a)). On the other hand, the fit
is perfect in the limit of the irreversible quench, in which the
potential is instantaneously switched to its new position. This
corresponds to the limit $u \rightarrow \infty$ (or $k t_f
\rightarrow 0$). One finds
\begin{equation}
I _n (t_f \to 0) =\frac{1}{2}kz^2_0= \langle W_{\rm{diss}} \rangle
(t_f \to 0),
\end{equation}
since the dissipated work is exactly equal  to the average work done
in instantaneously placing the particle in the shifted potential.

In Fig.~\ref{boundsIn}.(b) we plot $I_n$, for different values of
$kt_f$, as a  function of the number of measured points. Note that
the biggest jumps in $I_n$ occur from $n=1$ to $n=2$, after which
the bound quickly saturates and slowly approaches the total mean
dissipated work. The dominant term in the convergence of $I_n$ to
$I_{\infty}$ is easily obtained from Eq.~(\ref{In!}):
\begin{equation} I_{\infty}-I_n\simeq
\frac{z_0^2k^2t_f}{12n^2}.
\end{equation}

We expect that this type of convergence, $1/n^2$, is valid for
a continuous Markov process. There is a plausibility argument for this
asymptotic behavior. The chain rule of the relative entropy, Eq.
(\ref{chainrule}), implies
\begin{equation}\label{asympsum}
I_{\infty}-I_n= \sum_{i=0}^{n-1}  \left\langle \ln
\frac{\mathcal{P}(y_i|x_i,x_{i+1})}{\tilde \mathcal{
P}(y_i|x_i,x_{i+1})}\right\rangle,
\end{equation}
where $y_i(t)$ stands for the ``piece'' of the trajectory $x(t)$
with $t\in [t_i,t_i+1]$.  Under the conditions $x(t_i)=x_i$, each of
these pieces looks similar to the trajectories depicted in  Fig.
\ref{scheme} and becomes a pinned diffusion process, which can
be written as \cite{misawa,misawa2}:
\begin{eqnarray}
y_i(t) &=& \frac{(t_{i+1}-t)[x_i+X(t)-X(t_i)]}{\Delta t}
\nonumber\\ &+&
  \frac{(t-t_i)[x_{i+1}-X(t_{i+1})+X(t)]}{\Delta t},
\end{eqnarray}
where $X(t)$ is a process satisfying the same dynamics as $x(t)$ but
with no restrictions. Averaging the above equation and assuming that
$\langle X(t)\rangle$ is an analytical function of $t$, one has:
\begin{equation}
\langle y_i(t)\rangle = \frac{(t_{i+1}-t)x_i+(t-t_i)x_{i+1}}{\Delta
t} + O(\Delta t^2).
\end{equation}
Therefore, the averages of $\langle y_i(t)\rangle$ in the forward
and backward processes can only differ by terms of order $\Delta
t^2$.

On the other hand, the relative entropy between two Gaussian
distributions with the same dispersion $\sigma$ and averages $\mu_1$
and $\mu_2$ is given by: \begin{equation}\label{diffmean}
D(\rho_1||\rho_2)=\frac{(\mu_1-\mu_2)^2}{2\sigma^2}.
\end{equation}
Let us assume that $y_i(t)$ can be approximated by a Gaussian
process with dispersion of order $\sqrt{t}$. If the external
parameter does not affect the dispersion, the dominant term in the
relative entropies in Eq.~(\ref{asympsum}) will be given by the
difference between averages, i.e., by Eq.~(\ref{diffmean}):
\begin{equation}\left\langle \ln \frac{\mathcal{ P}(y_i|x_i,x_{i+1})}{\tilde
\mathcal{P}(y_i|x_i,x_{i+1})}\right\rangle \sim
\frac{(\mu_F-\mu_B)^2}{\sigma^2}\sim \Delta t^3.
\end{equation}
If this is the case, then the asymptotic approach to the exact work
is the same as in our example:
\begin{equation} I_{\infty}-I_n \sim n\Delta t^3 \sim \frac{1}{n^2}.
\end{equation}

\subsection{Underdamped Brownian particle in a suddenly quenched trap}\label{sec:under}

In  the previous example, we discussed the effect of  coarse
graining in time for the measurement of the single relevant variable
at hand, namely the position of the overdamped Brownian particle.
Now we address the additional question about the role of specific
variables (or degrees of freedom) in revealing the dissipation. For
the illustration of this point we naturally turn to underdamped
Brownian particles, where both position and momentum of the particle
are relevant. Instead of considering a moving harmonic potential
with fixed strength, we study another experimentally significant
scenario: a non-moving harmonic potential undergoing an
instantaneous quench in its stiffness, say at the initial time $t=0$
from a frequency $\omega_0$ to the frequency $\omega_1$. See the scheme in figure \ref{fig:scheme}.

The average work dissipated $\langle W_{\rm{diss}}\rangle$  in the
instantaneous quench can be evaluated as follows. The potential
energy of the particle when at a position $x$, is given by
$V_i(x)=m\omega_i^2x^2/2$, where $\omega_i$ is the harmonic
frequency, with $i=0$ and $i=1$ before and after the quench,
respectively. The probability distribution of the position at the
moment of the quench is given by
$\rho^{\rm{eq}}
_0(x)=\exp(-V_0(x)/T)/Z_0$ (as before, Boltzmann's
constant is absorbed in the temperature for simplicity of notation).
Here $Z_0$, the normalization constant, is the familiar partition
function. Averaging with respect to this distribution (notation
$\langle...\rangle_0$), we conclude that the average work associated
to the quench is given by $ \langle W\rangle=\langle
V_1(x)\rangle_0-\langle
V_0(x)\rangle_0=(T/2)(\omega_1^2/\omega_0^2-1)$. The corresponding
change in free energy is found to be $\Delta
F=-T\ln(Z_1/Z_0)=T\ln(\omega_1/\omega_0)$. Therefore, the total
dissipation in the irreversible instantaneous quench reads
\begin{equation}
\langle W_{\rm{diss}}\rangle \equiv \langle W\rangle -\Delta F=
\frac{T}{2}\left( \ln \frac{\omega_0^2}{\omega_1^2}+\frac{\omega_1^2}{\omega_0^2}-1
\right).
\end{equation}
Note that the total dissipated work is always positive  due to the irreversible nature of the process.

\begin{figure} [t]
\begin{center}
  \includegraphics[width=7.7cm]{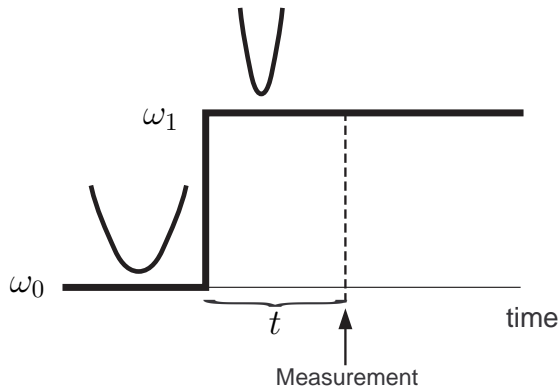}
  \caption{Schematic representation of the quenching experiment. The only information collected takes place at time $t$ after the quench.
  Notice that at any time $t$, in the backward process the system is in equilibrium at $\omega_1$, $\tilde\rho(t)=\rho^{\rm{eq}}
(\omega_1)$, whereas in the forward process the
  system is relaxing precisely towards this equilibrium state.
  }
\label{fig:scheme}
\end{center}
\end{figure}

As available statistical information we consider the probability
distribution $\mathcal{P}_{\rm{cg}}\equiv \rho$ for  position $x$
and momentum $p$ at a any single instant of time $t$ after the
quench (see Fig. \ref{fig:scheme}). As we have already shown in the
first example, statistical information at just one particular time
must provide a lower bound for the dissipated work corresponding to
such quench:
\begin{equation} \label{boundxp}
\langle W_{\rm{diss}} \rangle \geq
 k_B T D(\rho(x,p;t) ||{\tilde{\rho}}(x,-p;t))  \geq 0.
\end{equation}
 Below we will elucidate the effect of such
coarse-graining  implied in the punctual measurement in time (at
time $t$ after the quench) plus the effect of a reduction in the number of variables,
this is, measuring only $x$, only $p$ or both. We anticipate that
this will lead to inequalities such as
\begin{eqnarray} \label{boundx}
\langle W_{\rm{diss}} \rangle
 & \geq &
k_B T D(\rho(x;t) ||{\tilde{\rho}}(x;t)) \geq 0,\\
\langle W_{\rm{diss}} \rangle
& \geq &
 k_B T D(\rho(p;t) ||{\tilde{\rho}}(-p;t))\geq 0. \label{boundp}
\end{eqnarray}

To explicitly obtain the bounds from the coarse-grained relative entropies
appearing in equations (\ref{boundxp}), (\ref{boundx}) and (\ref{boundp}), we need to evaluate
the probability distributions in forward and backward scenario.
The derivation for the backward scenario is very simple;
the system starts at canonical equilibrium with frequency $\omega_1$ and the
quench is performed at the end of the experiment ($t=0$ in forward
time,  which is the final time in the reverse experiment).
The particle is then at canonical equilibrium with respect to the
frequency $\omega_1$ throughout the process:
\begin{equation}
\tilde\rho(x,p;t)=\rho_{1}^{\rm{eq}}
(x,p)=
\frac{\exp{[-(p^2/2m+m\omega_1^2x^2/2)/T}]}{Z(\omega_1)}.
\end{equation}
Note that the above distribution is even in $p$, namely $\tilde\rho(x,p;t)=\tilde\rho(x,-p;t)$,
and Gaussian with the following
moments:
\begin{eqnarray}\label{condin1}
 \langle \tilde x\rangle =
\langle \tilde p\rangle &=&
\langle \tilde{x p}\rangle = 0,  \nonumber \\
 \langle \tilde x^2 \rangle & = & T/(m\omega_1^2),  \nonumber \\
\langle \tilde p^2\rangle & = & mT.
\end{eqnarray}

One the other hand, in the forward scenario, the initial condition
is canonical with respect to the initial frequency $\omega_0$,
$\rho(x,p;0)=\rho_0^{\rm{eq}} (x,p)$. At $t=0$ the frequency is
suddenly changed to $\omega_1$ and then kept constant along the
whole process and, therefore, the evolution of the system in the
forward process consists of a relaxation to the new equilibrium
state, $\rho_1^{\rm{eq}} (x,p)$. Then we are free to decide what we
call the final time of the experiment  and hence the choice of the
measurement time after the quench is also completely free. We write
the familiar equations of motion for such underdamped Brownian
particle for $t>0$,
\begin{eqnarray} \label{le}
\dot p(t) &=&- m\omega_1^2
x(t)-\lambda p(t)/m +\xi(t), \nonumber \\
\label{le2}
\dot x(t) &= & p(t)/m,
\end{eqnarray}
where $\lambda$ is the friction coefficient, and $\xi$ is Gaussian
white noise with strength determined by the fluctuation dissipation
theorem, $\langle \xi(t)\xi(t')\rangle = 2\lambda T\delta(t-t')$.
The initial condition is stipulated by the fact that prior to the
quench at $t=0$, the system is at equilibrium in a harmonic
potential with strength $\omega_0$, i.e. it is bi-Gaussian with
(cf. Eq.~(\ref{condin1}))
\begin{eqnarray}\label{condin}
 \langle x\rangle_{(t=0)} =
\langle p\rangle_{(t=0)} & = &
\langle x p\rangle_{(t=0)}= 0, \nonumber \\
\langle x^2\rangle_{(t=0)} & = & T/(m\omega_0^2), \nonumber \\
 \langle p^2\rangle_{(t=0)} & = & mT.
\end{eqnarray}
Since the Langevin equation is linear, the resulting time dependent
probability distribution $\rho(x,p;t)$ remains a Gaussian.
Therefore, it is sufficient to evaluate the ensuing time evolution
of first- and second-order moments. Since there is no shift in the
center  position of the harmonic potential, the average position and
momentum stay equal to zero: $\langle x(t)\rangle =\langle
p(t)\rangle=0$. The second order moments obey the
following evolution equations which are easily obtained from the evolution equations (\ref{le}):
\begin{eqnarray}\label{eqs}
\frac{d}{dt} \langle x^2\rangle &=& \frac{2}{m}\langle xp\rangle, \nonumber\\
\frac{d}{dt} \langle xp\rangle &=& \frac{1}{m}\langle p^2\rangle -
m\omega_1^2 \langle x^2\rangle -\frac{\lambda}{m} \langle xp\rangle,\nonumber\\
\frac{d}{dt} \langle p^2\rangle &=& -2m\omega_1^2 \langle xp\rangle -\frac{2\lambda}{m}
\langle p^2\rangle +2\lambda T,
\end{eqnarray}
which have to be solved with the above mentioned initial conditions.
One finds:
\begin{eqnarray}  \label{eqx2}
\langle x^2\rangle _{t} & = & \frac{T}{mw_1^2}\left[
1-\frac{\omega}{1-\sigma^2}\; e^{-t\lambda/m}
\mathcal{C}
\right],\nonumber\\
\label{eqxt}
\langle xp\rangle_t & = & \frac{mT}{\lambda}
\frac{\omega}{1-\sigma^2}
e^{-t\lambda/m}\left[
1-\cosh (t \nu) - \frac{m}{\lambda}\nu\sinh (t\nu)
\right],\nonumber\\
\label{eqp2}
\langle p^2\rangle_t & = & mT \left[   1+
\frac{\sigma^2}{1-\sigma^2} \; \omega \;
e^{-t\lambda/m}
\sinh^2 \left( t \nu/2 \right)
\right],
\end{eqnarray}
where
\begin{eqnarray}
\mathcal{C}\equiv
\sigma^2/2-(1-\sigma^2/2)\cosh [t\nu]
-\frac{m}{\lambda}\nu\sinh[t\nu],\\
\omega \equiv \left( \frac{\omega_1}{\omega_0} \right)^2-1,
\; \;
\nu \equiv
\frac{\lambda}{m} \sqrt{1-\sigma^2},
\; \;
\sigma\equiv \frac{2m\omega_1}{\lambda}.
\end{eqnarray}
Note the switch from a monotonously decay ($\nu$ real) to an oscillatory one ($\nu$ imaginary)
of the above solutions for the moments as $\sigma $ crosses the value $1$ from below.

\begin{figure} [t]
\begin{center}
\includegraphics[angle=270, width=7.7cm]{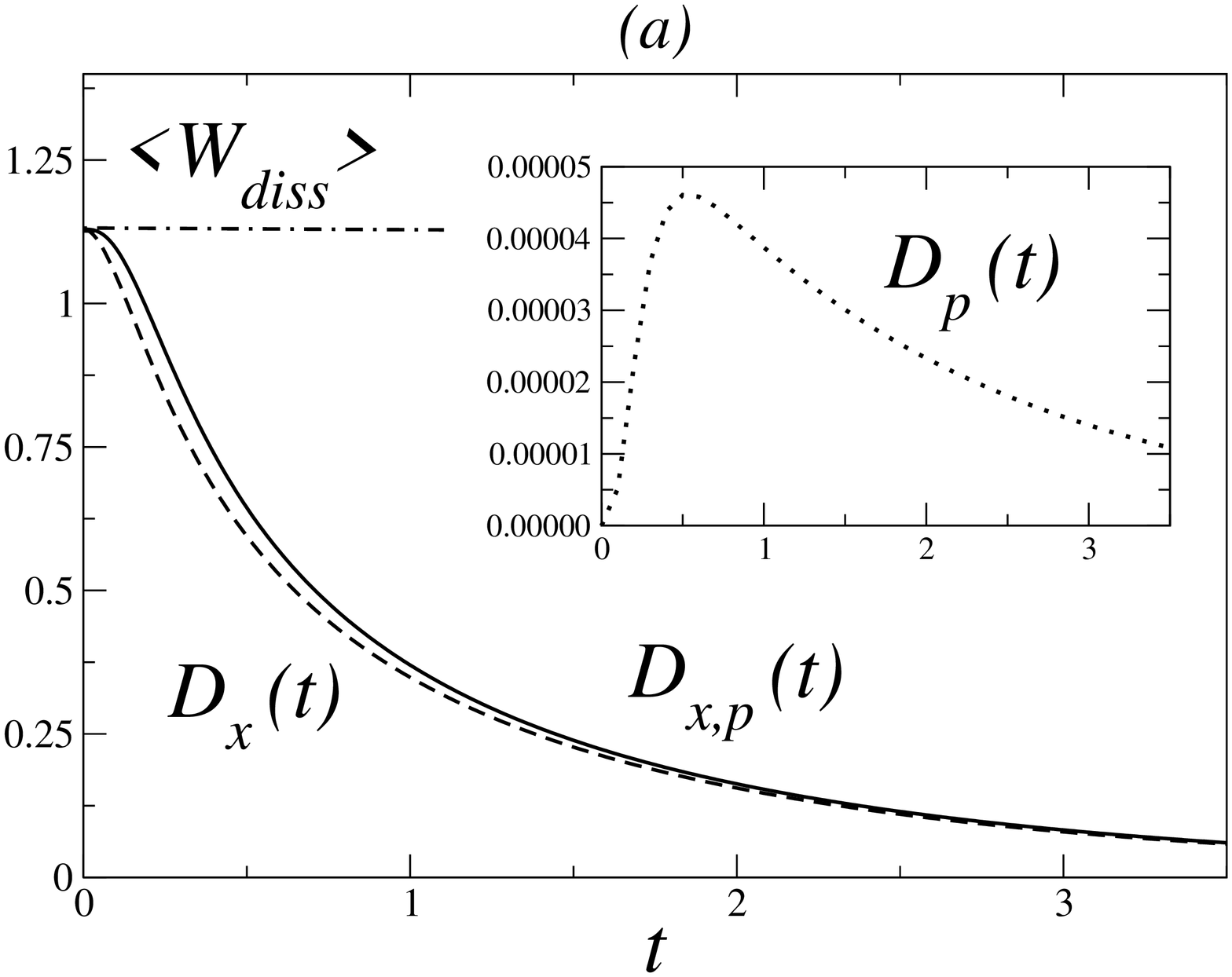}
\includegraphics[angle=270, width=7.7cm]{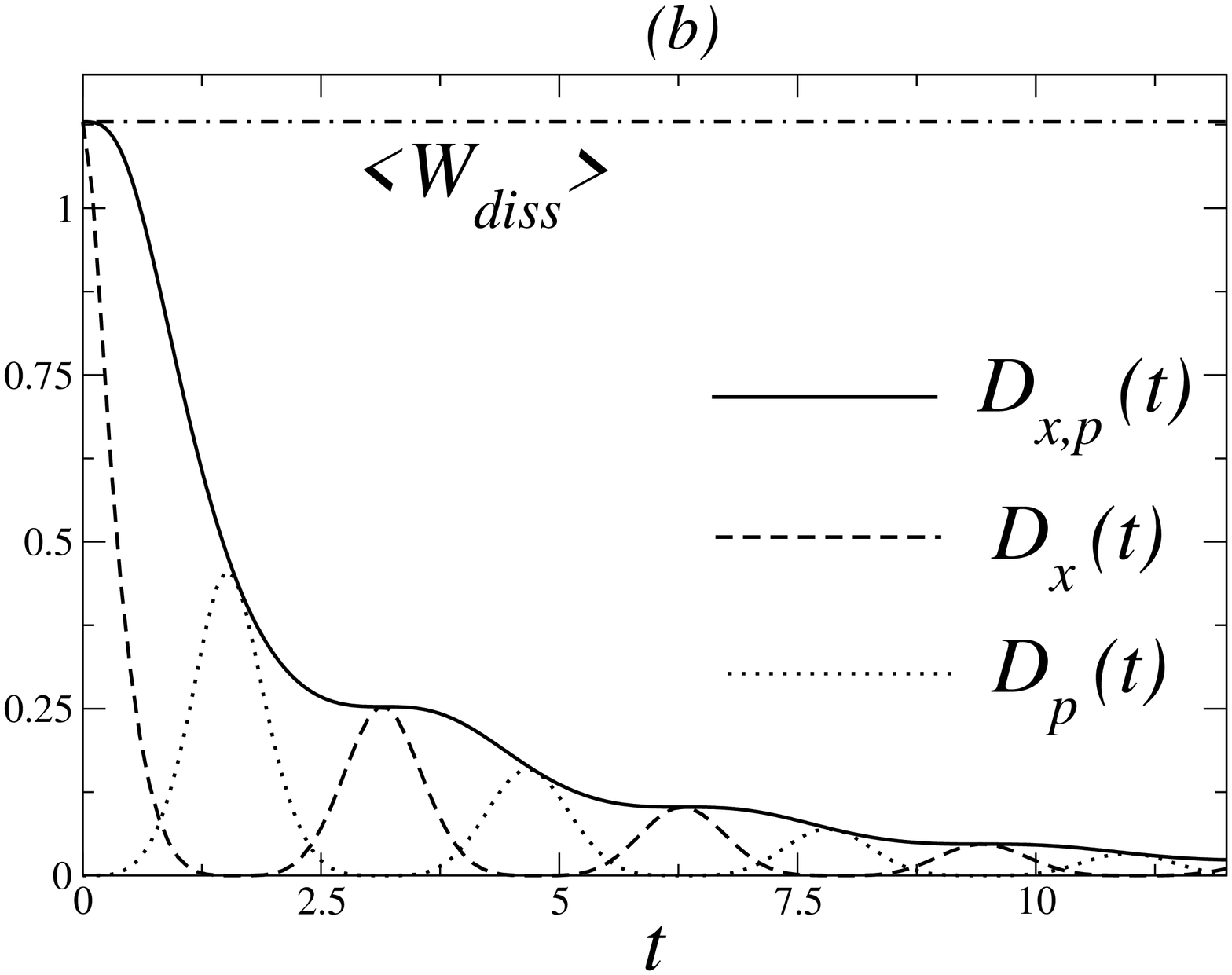}
\caption{(a) Relative entropies $D_{x,p}$, $D_x$ and $D_p$ measured at one single time $t$ after the quench. Strongly damped regime ($\sigma<1$): friction dominates inertia. $D_{x,p}$ and $D_{x}$ decay monotonically and almost coincide, while $D_p$ is very small and
goes through a rise-and-fall.
(b) Underdamped  regime ($\sigma>1$): inertia dominates friction,
resulting in an out of phase oscillatory decay of the relative entropies $D_x$ and $D_p$. Note that in both cases (a) and (b),  the position variable $x$
captures at $t=0$ the full information on dissipation, namely, $\langle W_{\rm{diss}}\rangle$. } \label{Dxpfig2}
\end{center}
\end{figure}

We are now in position to evaluate the relative entropy
between $\rho(x,p;t)$ and $\tilde\rho(x,-p;t)$, which in this case
can be considered as a distance between the relaxing time dependent distribution $\rho(x,p;t)$
and its final equilibrium state $\rho_1^{\rm{eq}}
(x,p)$, only
reached for $t\to \infty$.
Since both densities are Gaussian (and the
backward distribution is even in $p$), the following result
is obtained:
\begin{eqnarray} \label{DKL2}
D_{x,p}(t) & \equiv & D(\rho(x,p;t)||\tilde\rho(x,-p;t)) \nonumber \\
&= &
 \ln \sqrt{ \frac{{\rm det}\, \tilde C_2}{{\rm det}\,
C_2} } +\frac{{\rm Tr} (\tilde C_2^{-1}C_2)}{2}-1,
\end{eqnarray}
where $C_2$ and $\tilde C_2$ are the covariance matrices of the
forward and backward distributions, respectively
\begin{equation} C_2=\left(
\begin{array}{cc} \langle x^2\rangle_t & \langle xp\rangle_t \\
\langle xp\rangle_t & \langle p^2\rangle_t \end{array}\right),
\;\;\;\;
\tilde C_2=\left(
\begin{array}{cc} \langle \tilde x^2\rangle  & \langle \tilde{xp}\rangle\\
\langle \tilde{xp}\rangle & \langle \tilde p^2\rangle
\end{array}\right).
\end{equation}
The above result can be further simplified to
\begin{equation} \label{DKLxp}
D_{x,p}(t)= \frac{1}{2} \left( \ln \frac{\langle \tilde
x^2\rangle \langle \tilde p^2\rangle  }{\langle  x^2\rangle_t
\langle p^2\rangle_t-\langle  xp\rangle_t^2} + \frac{\langle
x^2\rangle_t}{\langle \tilde x^2\rangle}+ \frac{\langle
p^2\rangle_t}{\langle \tilde p^2\rangle} -2   \right).
\end{equation}
From now on, subindices in $D$ refer to the variables contained in the
probability distributions with which the relative entropy is evaluated.
So when momentum is integrated out from the probability distribution,
the relative entropy at time $t$ of the position distributions,
$D(\rho(x;t)||\tilde\rho(x;t))$, yields
\begin{equation} \label{DKLx}
D_x(t)
=\frac{1}{2}\left( \ln \frac{\langle \tilde x^2\rangle}{\langle  x^2\rangle_t}
 +\frac{\langle  x^2\rangle_t}{\langle \tilde x^2\rangle} -1   \right),
\end{equation}
and its momentum analog, $D(\rho(p;t)||\tilde\rho(-p;t))$, is
\begin{equation} \label{DKLp}
D_p(t)=
\frac{1}{2}\left( \ln \frac{\langle \tilde p^2\rangle}{\langle  p^2\rangle_t}
 +\frac{\langle  p^2\rangle_t}{\langle \tilde p^2\rangle} -1   \right).
\end{equation}

We insert in (\ref{DKLxp}), (\ref{DKLx}) and (\ref{DKLp}) the
expressions of the second moments calculated previously. With these
explicit results (depicted in figure \ref{Dxpfig2}), we can discuss
how well various relative entropies capture the information on
the dissipation. First we note  that at the moment of the quench
($t=0$), the statistics of the position variable can account for the
total work dissipated, $TD_x(0)=\langle W_{\rm{diss}}\rangle$,
 while no information is available from the momentum variable, $D_p(0)=0$. The reason is that the
 position at the time of quench is enough to evaluate the work
\cite{bounds}.

Secondly, it is known that the relative entropy between the
probability distribution of a Markov process and its corresponding
stationary state is a strictly decreasing function of time
\cite{cover}. Hence, $D_{x,p}(t)$ must be so, as one can check from
our  calculations plotted in figure \ref{Dxpfig2}. On the other
hand, when only one of the variables is taken into account, the
relative entropies can exhibit a richer phenomenology. The behavior
is rather different in the weakly damped regime than in the strongly
damped one. In the strongly damped case ($\sigma<1$) the relative
entropies $D_{x,p}(t)$ and $D_x(t)$ just decay monotonically with
time, see figure {\ref{Dxpfig2}.(a). However, we obtain a
non-monotonous behavior in the relative entropy of the momentum
distribution, which is explained as follows. The equilibrium
distribution of the momentum does not depend on the frequency of the
oscillator. Therefore, at the quench time, the forward and backward
momentum distributions are identical. However, once the potential is
quenched, the  momentum distribution will depart from equilibrium,
due to transfers from potential to kinetic energy, to relax back to
the same distribution at a later time. As a consequence $D_p(t)$
increases from $D_p(0)=0$, reaches a maximum and decays back to zero
for long time, as can be seen in the inset of figure
\ref{Dxpfig2}.(a). The maximum is however very low, since damping is
strong.

We can see a more pronounced and interesting effect in the
underdamped case ($\sigma>1$). The main  results are represented in
figure \ref{Dxpfig2}.(b).  Note the oscillatory exchange  of
information on dissipation between the position and velocity
variables and the decay of the total information contained in
$D_{x,p}(t)$. The behavior of $D_x(t)$ and $D_p(t)$ is induced by
oscillations in the  potential and kinetic energy of the particle.
The relative entropy of the $x$ distributions can be written as
(cfr. Eq.~(\ref{DKLx})):
\begin{equation} D_{x}(t)=\frac{\alpha_V(t)-1- \ln
\alpha_V(t)}{2}
\end{equation}
where $\alpha_V(t)$ is the ratio between the potential energy at
time $t$ and the equilibrium potential energy (at frequency
$\omega_1$). A similar expression can be obtained for $D_p(t)$ from
Eq.~(\ref{DKLp}). Therefore, the relative entropy of $x$ and $p$ can
be considered as a measure of the departure of the potential and
kinetic energy, respectively, from their equilibrium values. These
energies turn to oscillate at  frequency $\nu$, twice the
characteristic frequency of the damped oscillator.

The  behavior of $D_{x,p}(t)$ observed in Figure \ref{Dxpfig2}.(b)
can be better understood by rewriting the relative entropy in
(\ref{DKLxp}) as follows:
 \begin{eqnarray} \label{DDD}
D_{x,p}(t)=D_x(t)+D_p(t) +\frac{1}{2}\ln \left( \frac{1}{1-r_t} \right),
\end{eqnarray}
where the correlation coefficient $r_t$ is given by
\begin{equation}
r_t \equiv \frac{\langle xp \rangle_t^2}{\langle x^2\rangle_t \langle p^2 \rangle_t}.
\end{equation}
Since $0\leq r_t \leq 1$, we first note that  the last term in the
r.h.s of equation (\ref{DDD}) is always positive, hence:
\begin{equation}\label{Dge}
D_{x,p}(t) \geq D_x(t)+D_p(t).
\end{equation}
Therefore, in the present case, the sum of information on the
dissipation gathered separately from position and momentum is
smaller than that from both variables taken together. The equality
sign in (\ref{Dge}) is realized  if $r_{t}=0$, this is, when
$\langle xp \rangle_{t}=0$. From the oscillating analogue of the
expression for $\langle xp\rangle_{t}$ in (\ref{eqxt}), one easily
verifies that this occurs at specific times $t=\frac{2\pi n}{\tilde
\nu}$,  where $\tilde \nu= \frac{\lambda}{m}\sqrt{\sigma^2-1}$.
Since the variables are Gaussian,  the condition of zero correlation
is tantamount to the independency of position and momentum.

Another interesting feature observed in  figure \ref{Dxpfig2} is
that one of the variables, either $x$ or $p$, loses all information
on dissipation at  another set of specific times. From equations
(\ref{DKLx}) and (\ref{DKLp}) one finds that this occurs if $\langle
x^2 \rangle_t =\langle \tilde x^2 \rangle$ or $\langle p^2 \rangle_t
=\langle \tilde p^2 \rangle$ respectively. This is in agreement with
the more general observation that the relative entropy of a specific
degree of freedom is zero when, at a given time, the detailed
balance condition holds, namely, when at that time the forward and
backward distributions are equal.

On the whole, an intricate transfer of information on dissipation is
taking   place between position and momentum of the underdamped
Brownian particle. At the same time, such information on the total
mean dissipated work is irreversibly lost by the punctual (one-time)
relative entropy of $x$ and $p$ and transfered to the heat bath
variables as time goes by.

\subsection{Flow of information between coupled oscillators. }\label{sec:under2}

\begin{figure} [t]
\begin{center}
  \includegraphics[angle=0, width=7.5cm]{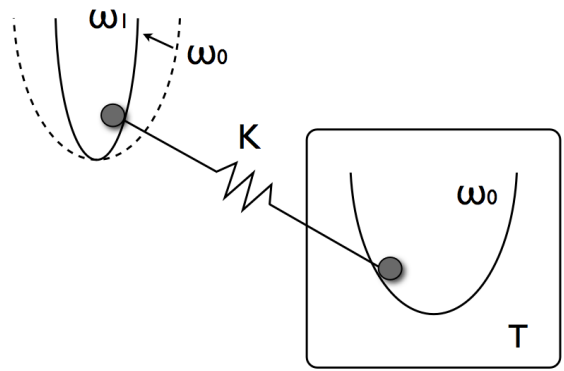}
  \caption{Scheme of the last study case: the information
  on dissipation due to a sudden quench in the stiffness
  of the potential from $\omega_0$ to $\omega_1$ in one
  subsystem is transmitted, through a linear coupling of
  strength $K$, to a second subsystem immersed in a heat
  bath at temperature $T$, where such information will get irreversibly lost.
 }
\label{flowmodel}
\end{center}
\end{figure}

To complete the picture, we next consider the case of a harmonically
bound underdamped Brownian particle that is indirectly in contact (via a second Brownian
particle) to a heat bath. The idea is that, by monitoring this second
particle, we are including some information on the heat bath, of
which it is supposed to be part. The Langevin equations of motion
that describe the system read:
\begin{eqnarray}
 m\ddot x_1 & = & -m\omega^2(t) x_1 -K(x_1-x_2),\\
 m\ddot x_2 & = & -m\omega_0^2 x_2-K(x_2-x_1)+\xi(t) - \lambda \dot x_2 ,
\end{eqnarray}
where $\langle \xi(t)\xi(t')\rangle = 2\lambda  T\delta(t-t')$. As
in the previous case, we consider the quench experiment: oscillator
$1$ is initially prepared in equilibrium with $\omega_0$. At $t=0$
we perform an instantaneous quench switching so that
$\omega(t)=\omega_1$ for $t>0$. Oscillator $2$ is kept throughout
at the same frequency $w_0$, linearly coupled  to oscillator $1$
with a strength $K$ and  immersed in the heat bath, which is modeled
by means of a fluctuating force $\xi(t)$ and a friction term
proportional to $\lambda$. See figure \ref{flowmodel}.

The time dependent probability distribution that characterizes the evolution of the whole system is a Gaussian, whose
first moments are simply
\begin{equation}
\langle x_1(t)\rangle =\langle p_1(t)\rangle=\langle x_2(t)\rangle =\langle p_2(t)\rangle=0.
\end{equation}
Thus we need to evaluate the second moments which,
having defined $K_0\equiv K+mw_0^2$ and $K_1 \equiv K+mw_1^2$,
obey the following set of evolution equations:
\begin{widetext}
\begin{equation}\label{fw4}
 \frac{d }{dt} \left( \begin{array}{c}
\langle x_1^2 \rangle \\
\langle p_1^2 \rangle \\
\langle x_1 p_1 \rangle \\
\langle x_2^2 \rangle \\
\langle p_2^2 \rangle \\
\langle x_2 p_2 \rangle \\
\langle  x_1 x_2\rangle \\
\langle x_1 p_2 \rangle \\
\langle x_2 p_1 \rangle \\
\langle p_1 p_2 \rangle
\end{array}
\right) =
 \left(
\begin{array}{cccccccccc}
0 &  0 & 2/m & 0 & 0 & 0 & 0 & 0 & 0 & 0 \\
0 &  0 & -2K_1 & 0 & 0 & 0 & 0 & 0 & 2K & 0 \\
-K_1 &  1/m & 0 & 0 & 0 & 0 & K & 0 & 0 & 0 \\
0 &  0 & 0 & 0 & 0 & 2/m & 0 & 0 & 0 & 0 \\
0 &  0 & 0 & 0 & -2\lambda/m & -2 K_0 & 0 & 2K & 0 & 0 \\
0 &  0 & 0 & -K_0 & 1/m & -\lambda/m & K & 0 & 0 & 0 \\
0 &  0 & 0 & 0 & 0 & 0 & 0 & 1/m & 1/m & 0 \\
K &  0 & 0 & 0 & 0 & 0 & -K_0 & -\lambda/m & 0 & 1/m \\
0 &  0 & 0 & K & 0 & 0 & -K_1 & 0 & 0 & 1/m \\
0 &  0 & K & 0 & 0 & K & 0 & -K_1 & -K_0 & -\lambda/m
\end{array}
\right)
\left( \begin{array}{c}
\langle x_1^2 \rangle \\
\langle p_1^2 \rangle \\
\langle x_1 p_1 \rangle \\
\langle x_2^2 \rangle \\
\langle p_2^2 \rangle \\
\langle x_2 p_2 \rangle \\
\langle  x_1 x_2\rangle \\
\langle x_1 p_2 \rangle \\
\langle x_2 p_1 \rangle \\
\langle p_1 p_2 \rangle
\end{array}
\right)
+\left( \begin{array}{c}
0 \\
0 \\
0 \\
0 \\
2 \lambda T \\
0 \\
0 \\
0 \\
0 \\
0
\end{array}
\right).
\end{equation}
\end{widetext}
The system can be solved explicitly using the appropriate initial
conditions corresponding to the equilibrium ensemble.
However, the analytic expressions are extremely
lengthy. In what follows, we will illustrate the obtained behavior via figures.

Since the joint distribution is Gaussian (and the backwards density
even in $p$), the relative entropy involving all four variables
$x_1,p_1,x_2$ and $p_2$ can be compactly expressed in terms of the
covariance matrices $C_4$ and $\tilde C_4$:
\begin{equation} \label{DKL4}
D_{x_1,p_1,x_2,p_2}(t)=
\frac{1}{2} \left[ \ln\left( \frac{{\rm det}\, \tilde C_4}{{\rm det}\,
C_4}\right) +{\rm Tr} (\tilde C_4^{-1}C_4)-4\right].
\end{equation}
The latter  are the following four-by-four symmetric matrices
\begin{equation} C_4=\left(
\begin{array}{cccc}
\langle x_1^2\rangle_t & \langle x_1p_1\rangle_t & \langle x_1x_2\rangle_t & \langle  x_1 p_2\rangle_t\\
\langle x_1 p_1\rangle_t & \langle p_1^2\rangle_t & \langle p_1 x_2\rangle_t & \langle p_1 p_2\rangle_t\\
\langle x_1 x_2\rangle_t & \langle p_1 x_2\rangle_t & \langle x_2^2\rangle_t & \langle x_2 p_2\rangle_t\\
\langle x_1 p_2\rangle_t & \langle p_1 p_2\rangle_t & \langle x_2 p_2\rangle_t & \langle p_2^2\rangle_t
\end{array}\right),
\end{equation}
and for the covariance matrix corresponding to the backwards excursion we explicitly find
\begin{equation} \tilde C_4=\left(
\begin{array}{cccc}
\frac{K_0 T}{Kmw_1^2+K_1 mw_0^2}
 & 0 & \frac{ K T}{Kmw_1^2+K_1 mw_0^2}
  & 0 \\
0 & mT & 0 & 0 \\
\frac{ K T}{Kmw_1^2+K_1 mw_0^2} & 0 & \frac{ K_1 T}{Kmw_1^2+K_1 mw_0^2} & 0 \\
0 & 0 & 0 & mT
\end{array}\right).
\end{equation}

From the above results, we can derive the relative entropy of all
available degrees of freedom of the system (both positions $x_1$ and
$x_2$, and momenta $p_1$ and $p_2$), along the whole time track with
a single time measurement. Similarly to the underdamped oscillator
case of the previous section, one can explore the behavior of the
relative entropies of all possible combinations of the $4$ degrees
of freedom. Some of them are plotted in figure \ref{D4x4fig}. All
the features of the model in the previous section are found here
too: the relative entropy of the whole system,
$D_{x_1,p_1,x_2,p_2}(t)$, decays monotonically in time and it is an
upper bound with respect to any other relative entropy accounting
for less degrees of freedom. The relative entropy of subsystem $1$,
$D_{x_1,p_1}(t)$,  is oscillating in an intrincate manner below the
former, together with $D_{x_1}(t)$ and $D_{p_1}(t)$, which transfer
information periodically and are modulated by $D_{x_1,p_1}(t)$.
Again, only when performed at the moment  of the quench does the
measurement concerning the position $x_1$ contain full information.

\begin{figure} [t]
\begin{center}
  \includegraphics[angle=270, width=7.7cm]{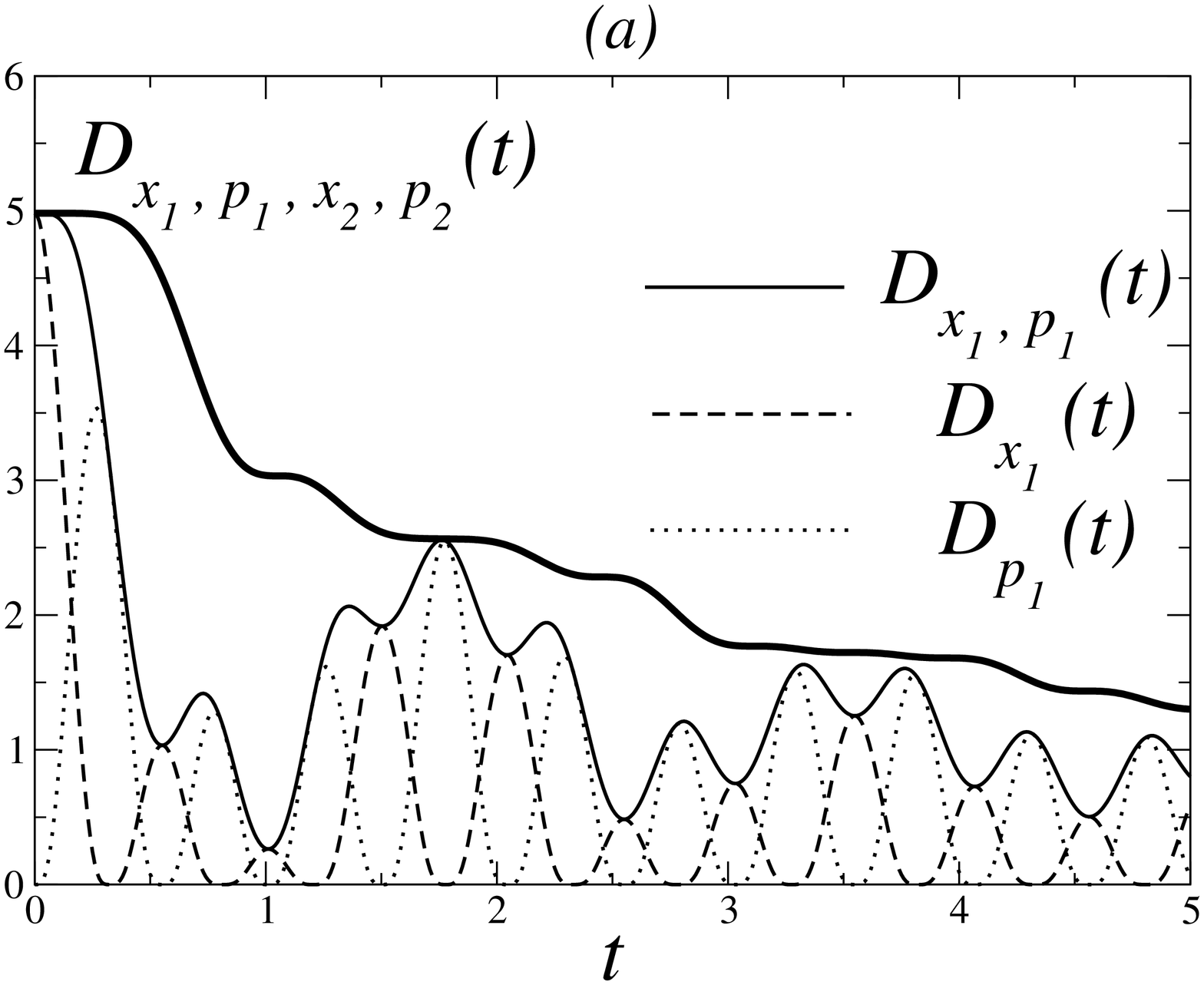}
 \includegraphics[angle=270, width=7.7cm]{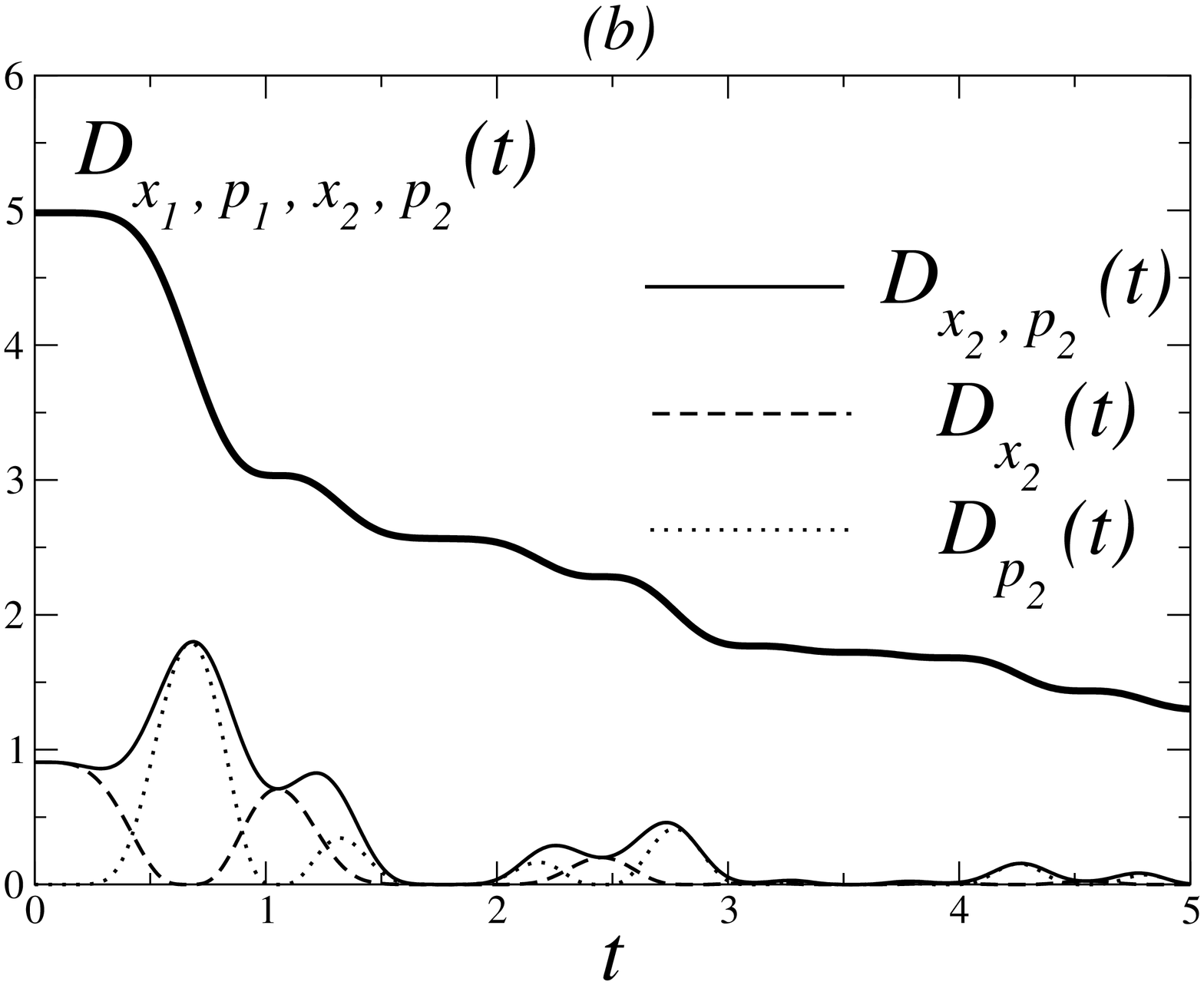}
  \caption{(a) Complex behavior of the relative entropies as a
  function of time for oscillator one. See text for discussion.
    (b) Same picture for oscillator two.
 }
\label{D4x4fig}
\end{center}
\end{figure}

The novelty in this case is that oscillator $1$ is not directly in
contact with the heat bath, but indirectly through oscillator $2$.
This allows for a more detailed study of  how the information on
dissipation contained in the first subsystem leaks out irreversibly.
First it has to flow to the second subsystem and then it is dumped
into the heat bath variables, where it is forever lost. Comparing
figure \ref{D4x4fig}.(a) with \ref{D4x4fig}.(b) we see that the
relative entropies of oscillator $2$ are significantly smaller than
those of oscillator $1$. The former receives information on the
dissipated work at the quench only through its coupling to $1$ and,
while it bounces back some of this information to $1$, its relative
entropy $D_{x_2,p_2}(t)$  decays much faster because it is directly
connected to the heat bath.

\begin{figure} [t]
\begin{center}
\includegraphics[angle=270, width=7.7cm]{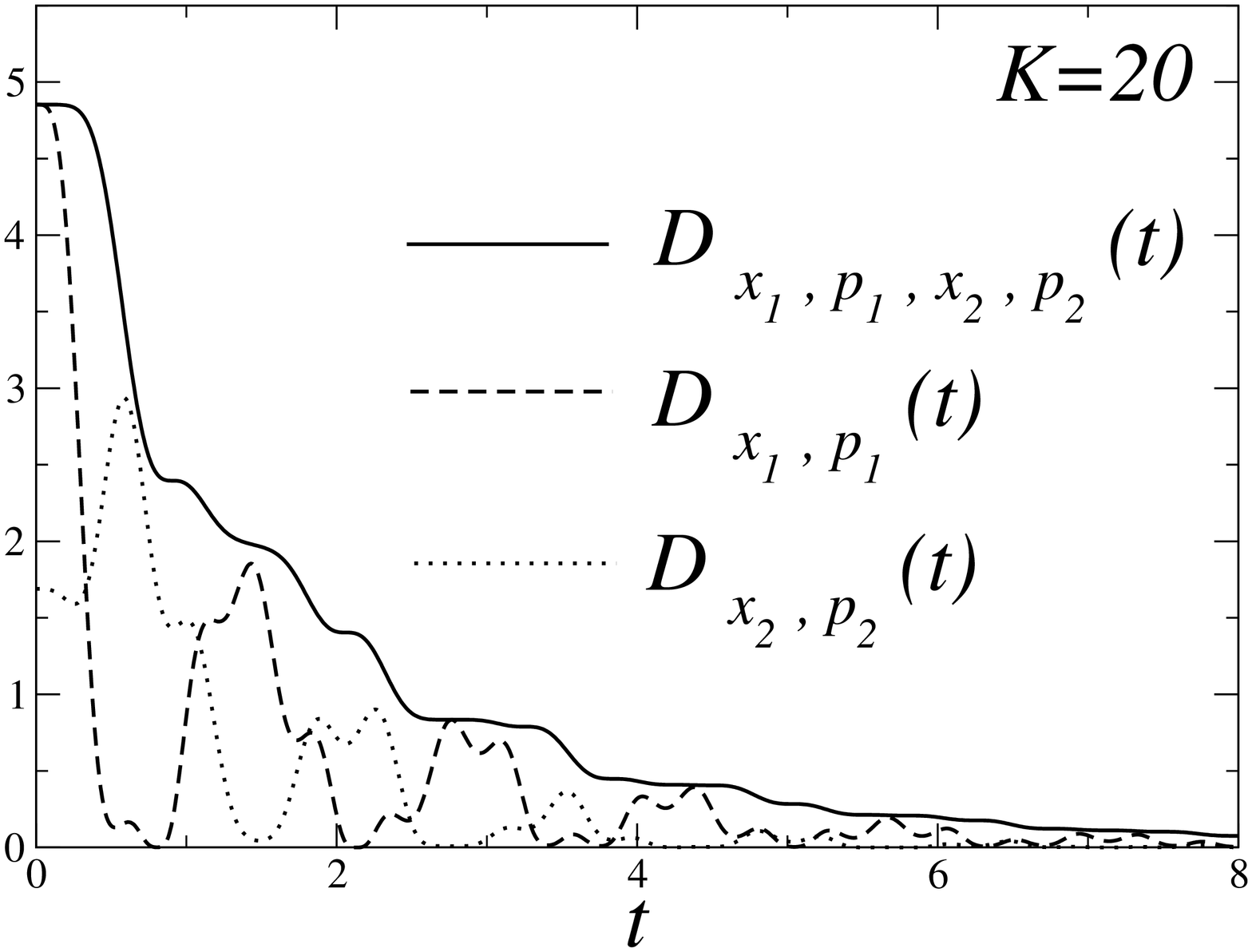}
\includegraphics[angle=270, width=7.7cm]{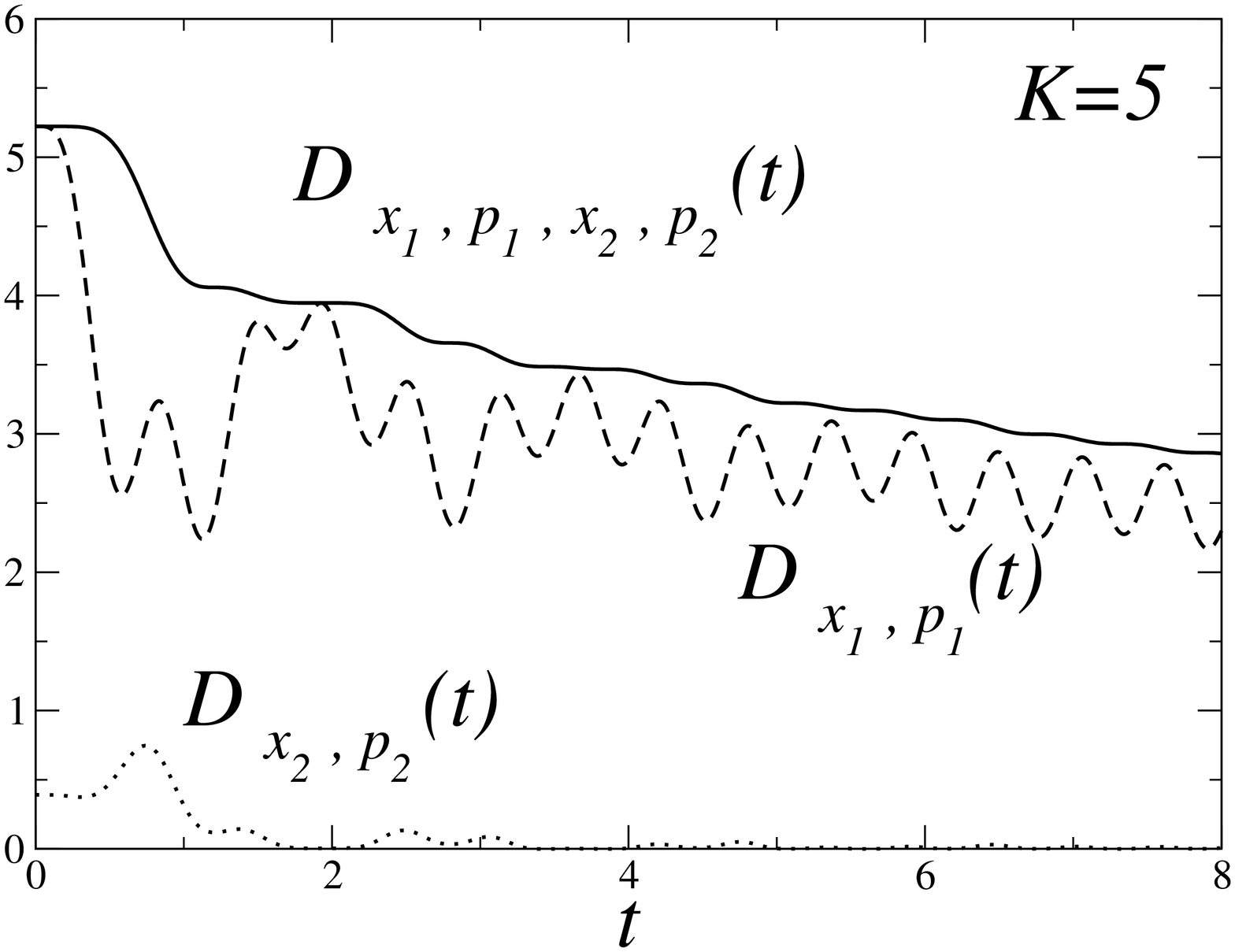}
 \caption{ Time evolution of the relative entropies of the whole system,
 $D_{x_1,p_1,x_2,p_2}(t)$, oscillator one, $D_{x_1,p_1}(t)$, and oscillator two, $D_{x_2,p_2}(t)$. In each plot a different coupling constant $K$ between the oscillators is used.
    }
\label{Dkexplore}
\end{center}
\end{figure}

The plateau for the relative entropy  appearing at short times in
figure \ref{D4x4fig} shows that most of the effect of the
dissipative process (the irreversible quench) still resides inside
the system formed by the two particles. In fact, oscillator $1$
keeps much of this information  while slowly transferring it to
oscillator $2$. This depends on the coupling constant $K$ hat
connects both oscillators. This dependence is illustrated in figure
\ref{Dkexplore}. For $K=20$ the general decay is fast since both
subsystems are well coupled and information on the quench can
quickly flow to the heat bath. However, for $K=5$ such flow is
reduced and the relative entropy of oscillator $2$ is almost zero
but yet there is a considerable difference between
$D_{x_1,p_1,x_2,p_2}(t)$ and $D_{x_1,p_1}(t)$. Therefore, while
oscillator $2$ is ``close to equilibrium", its correlation with
oscillator $1$ still carries relevant information on the
irreversible quench.
This conclusion is valid both for positions and momenta, separately, as seen
in Fig. \ref{figN} for $K=5$. As in the case of a single oscillator, information
flows from positions to momenta, but is mainly kept by the quenched
oscillator or by correlations with oscillator 2, which rapidly reaches
equilibrium.

\begin{figure*} 
\begin{center}
\includegraphics[angle=270, width=5.9cm]{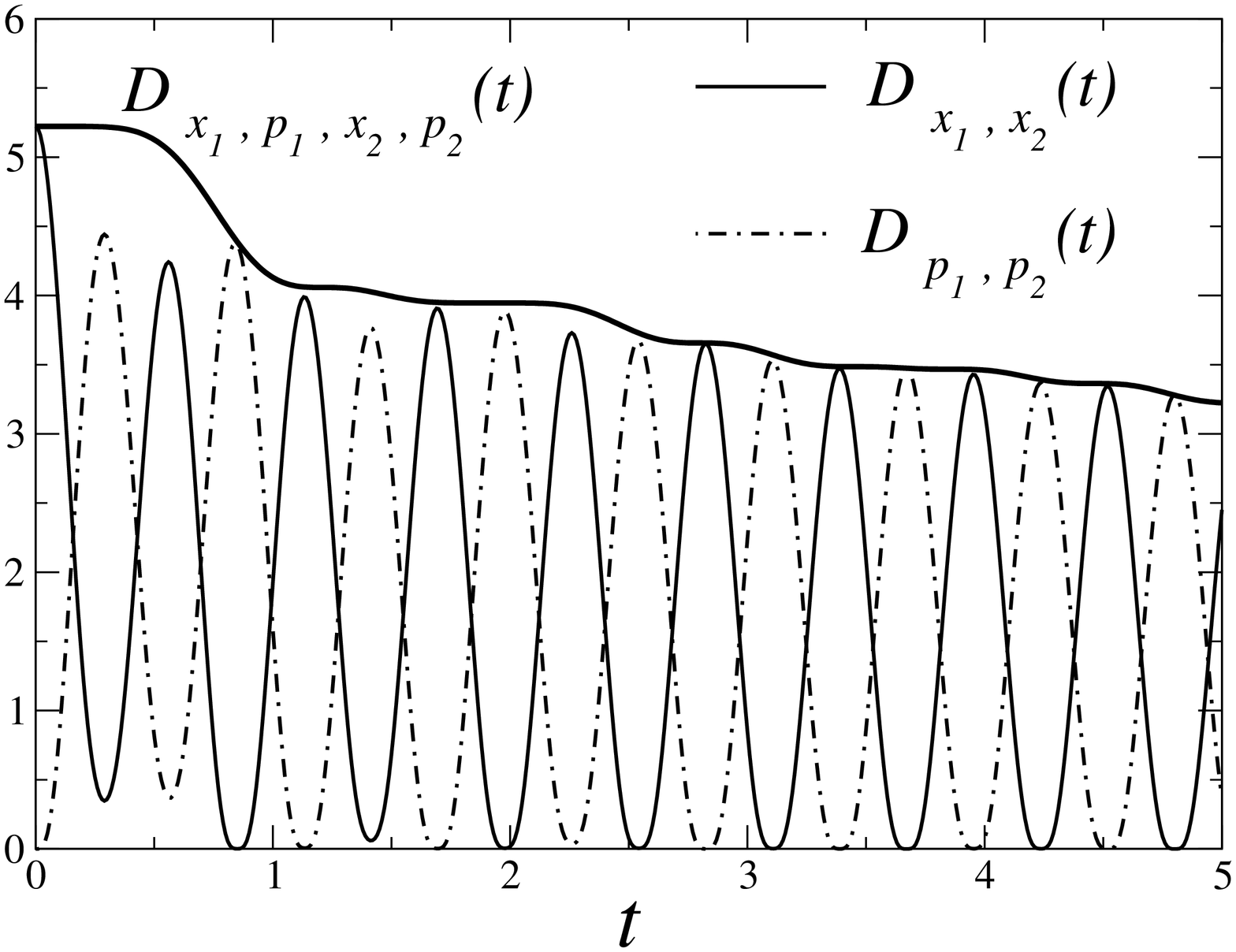}
\includegraphics[angle=270, width=5.9cm]{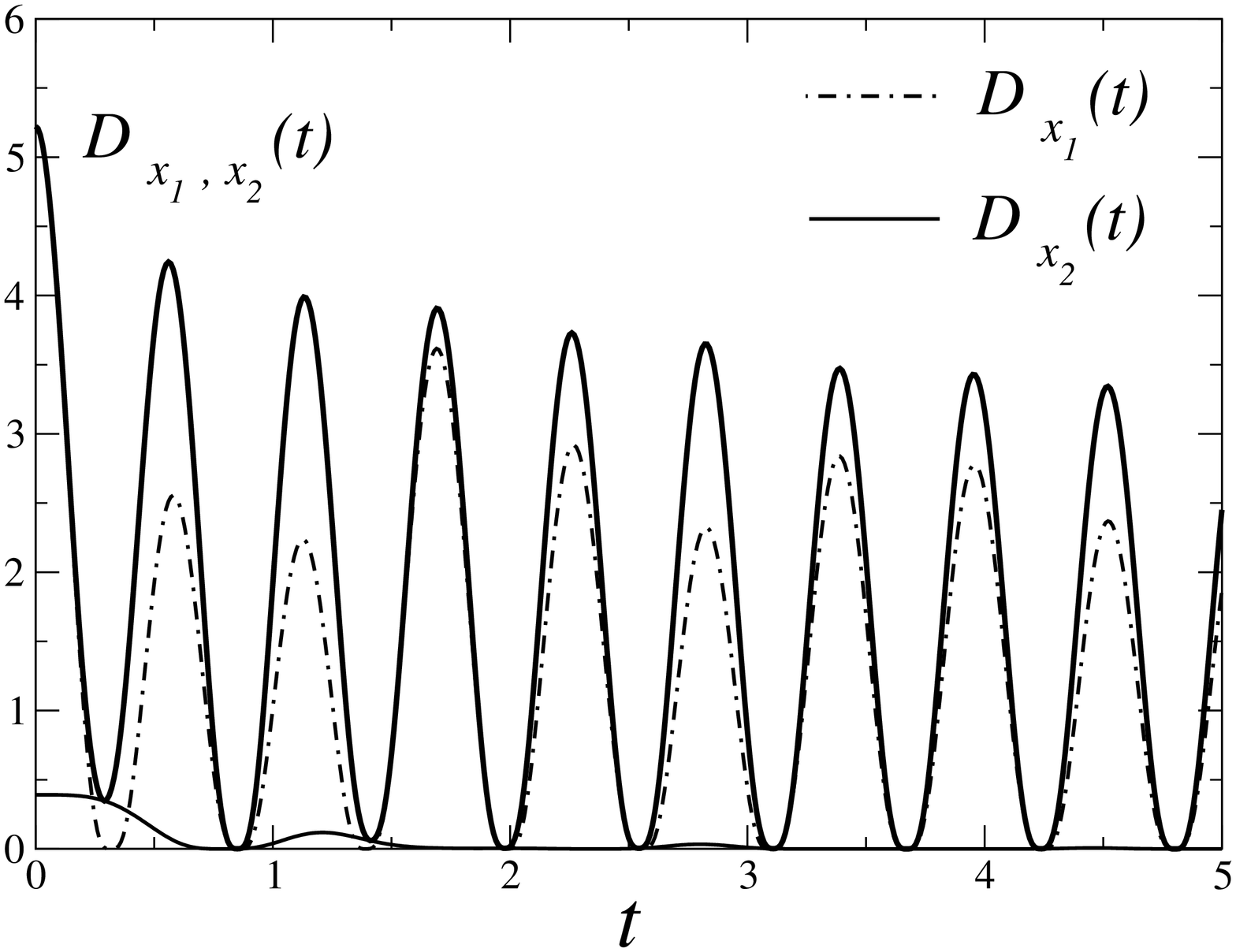}
\includegraphics[angle=270, width=5.9cm]{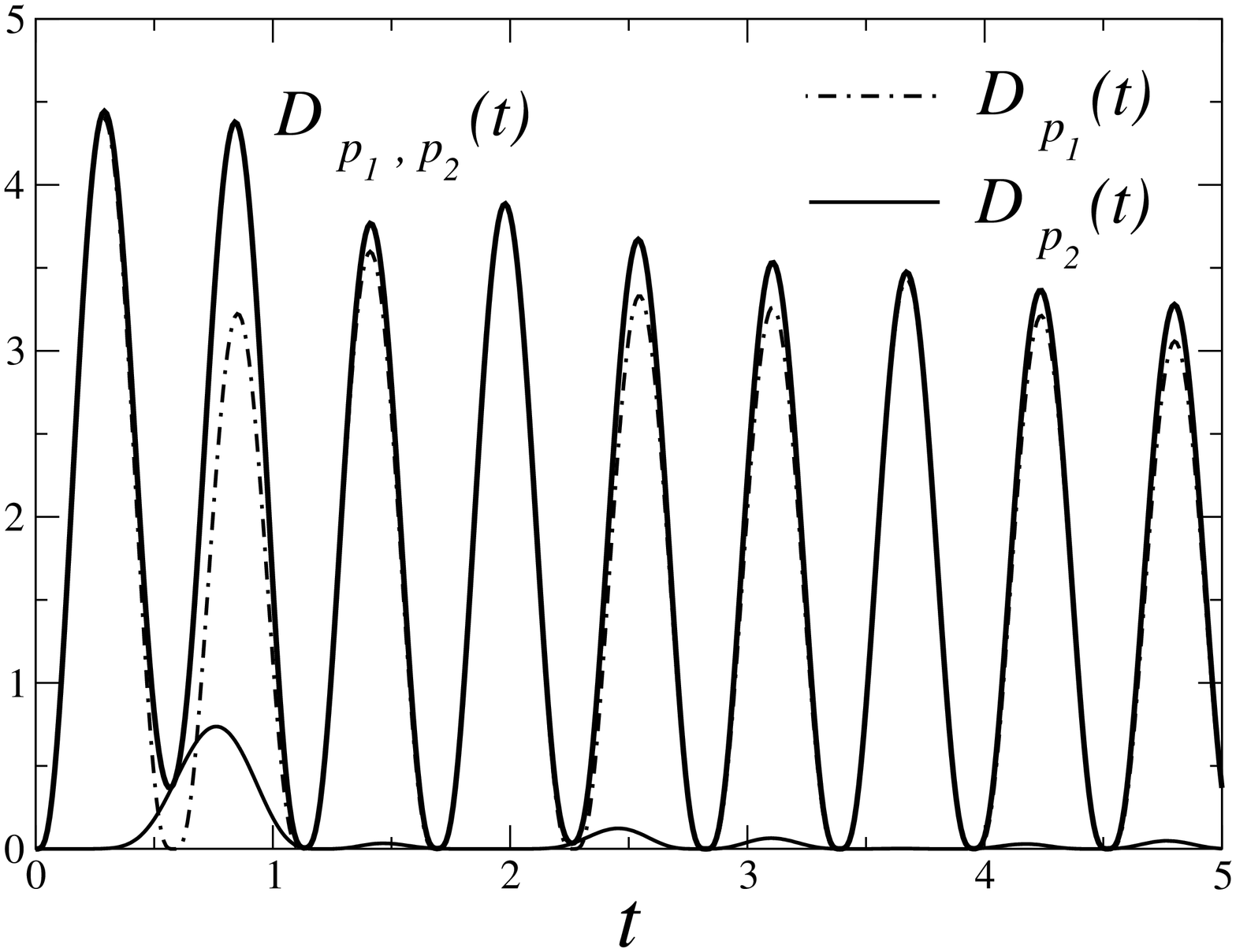}
 \caption{ Time evolution of several relative entropies illustrating the effect on correlations between positions and momenta.}
\label{figN}
\end{center}
\end{figure*}

\section{Conclusions}

Dissipation is related to our ability to  distinguish the arrow of
time. As was anticipated in earlier work in the literature, we find
that dissipation is proportional to the relative entropy between the
probability distributions of forward and backward excursions,
respectively. We have exploited such expression in terms of relative
entropy to give rise to lower bounds for the dissipation if only
partial information on the trajectories is available.

Several scenarios have been  discussed in order to illustrate how
dissipation can be bounded from below on the basis of reduced
information. First, when coarse-graining the continuous
trajectory of the system into a reduced finite number of
measurements, our analysis has shown that the resulting relative
entropy provides reasonably accurate bounds for the dissipation,
even with only a small number of intermediate measurement points. As
a generalization of our findings in this specific example, we
conjecture that the relative entropy obtained from $n$ measurements
approaches the exact value of the dissipation as $1/n^2$, for $n$
large. These results could be especially useful in real experiments
where trajectories are recorded at finite sampling rates.

Second, we have analyzed the effect of considering a subset of
variables instead of a detailed description of the system in a
quench process. In this case, the time-arrow information,
concentrated in the single position variable immediately after the
quench, is subsequently transferred  to the thermal bath and the
other variables. Of special interest is the case of two oscillators,
the first one undergoing a quench of its frequency and the second
one in contact with a thermal bath. We recall that the relative
entropy is constant when all the degrees of freedom of a Hamiltonian
system are taken into account \cite{diss}. One could then expect
that the information contained in the first oscillator would be
transferred to the second one before getting lost in the thermal
bath. However, our analysis calls into question this naive picture,
as we have shown that the oscillator coupled to the thermal bath is
the first to thermalize. The information on the dissipation is
mainly kept by the first oscillator or by correlations between the
two. The generalization of our analysis to long chains of
oscillators will help to further elucidate how information is spread
along many degrees of freedom.

 We acknowledge financial help from the Ministerio de Educacion
y Ciencia (Spain) under Grants FPU-AP-2004-0770 (A. G--M.) and
MOSAICO (JMRP), and from the FWO Vlaanderen.


\begin{thebibliography}{99}

\bibitem{onsager}
L. Onsager, Phys. Rev. {\bf 37}, 405 (1931).
\bibitem{onsager2}
L. Onsager, Phys. Rev. {\bf 38}, 2265 (1932).

\bibitem{prigogine}
I. Prigogine, {\it  Etude Thermodynamique des Ph\'enom\`enes
Irr\'eversibles} ( Desoer, Li\`ege) 1947.
\bibitem{prigogine2}
S. R. de Groot and P. Mazur,
{\it Non-Equilibrium Thermodynamics} (Dover, New York) 1984.

\bibitem{fluctuation}
D. J. Evans, E. G. D. Cohen, and G. P. Morriss, Phys. Rev.
Lett. {\bf 71}, 2401 (1993).
\bibitem{fluctuation2}
 G. Gallavotti and E. G. D. Cohen,  J. Stat. Phys. {\bf 80}, 931 (1995).
\bibitem{fluctuation3}
J. Kurchan, J. Phys. A {\bf 31}, 3719  (1998).
\bibitem{fluctuation4}
J. L. Lebowitz and H. Spohn,  J. Stat. Phys. {\bf 95}, 333 (1999).
\bibitem{fluctuation5}
 C. Maes,  J. Stat. Phys. {\bf 95}, 367 (1999).

\bibitem{work}
 G. N. Bochkov and Yu. E. Kuzovlev,  Zh. Eksp. Teor. Fiz. {\bf 72}, 238  (1977).
\bibitem{worka}
 G. N. Bochkov and Yu. E. Kuzovlev,  Sov. Phys. JETP {\bf 45}, 125 (1977).
\bibitem{workb}
 G. N. Bochkov and Yu. E. Kuzovlev,  Physica A {\bf 106}, 443  (1981).
\bibitem{workc}
 G. N. Bochkov and Yu. E. Kuzovlev,   Physica A {\bf 106}, 480  (1981).

\bibitem{work2}
 C. Jarzynski, Phys. Rev. Lett. {\bf 78}, 2690 (1997).
\bibitem{work2a}
 C. Jarzynski,  Phys. Rev. E {\bf 56}, 5018 (1997).
\bibitem{work3}
G. E. Crooks,  Phys. Rev. E {\bf 60}, 2721 (1999).
\bibitem{work4}
T. Hatano and S. I. Sasa,  Phys. Rev. Lett. {\bf 86}, 3463 (2001).
\bibitem{work5}
 B. Cleuren, C. Van den Broeck and R. Kawai,  Phys. Rev. Lett. {\bf 96}, 050601 (2006).

\bibitem{crooks}
G. E. Crooks,  J. Stat. Phys. {\bf 90}, 1481 (1998).

\bibitem{maes}
C. Maes and K. Netoc\"yny,  J. Stat. Phys. {\bf 110},  269 (2003).

\bibitem{seifert2005}
U. Seifert,  Phys. Rev. Lett. {\bf 95}, 040602 (2005).

\bibitem{gaspard2004}
P. Gaspard, J. Stat. Phys. {\bf 117}, 599 (2004).

\bibitem{luo}
Jiu-Li Luo, C. Van den Broeck and G. Nicolis,  Z. Phys. B {\bf 56}, 165 (1984).

\bibitem{schnakenberg}
J. Schnakenberg,  Rev. Mod. Phys. {\bf 48}, 571 (1976).

\bibitem{diss}
R. Kawai, J. M. R. Parrondo and C. Van den Broeck,  Phys. Rev. Lett. {\bf 98}, 080602 (2007).

\bibitem{bounds}
A. Gomez-Marin, J. M. R. Parrondo and C. Van den Broeck,
EPL, {\bf 82} 50002 (2008).

\bibitem{jar06}
C. Jarzynski,  Phys. Rev. E {\bf 73}, 046105 (2006).

\bibitem{jar07}
S. Rahav and C. Jarzynski,  J. Stat. Mech.: Theory Exp. (2007) P09012.

\bibitem{bly08}
R. A. Blythe,  Phys. Rev. Lett. {\bf 100}, 010601 (2008).

\bibitem{gaspard2007}
D. Andrieux, P. Gaspard, S. Ciliberto, N. Garnier, S. Joubaud and A. Petrosyan,
 Phys. Rev. Lett. {\bf 98}, 150601 (2007).

\bibitem{cover}
 T. M. Cover and J. A. Thomas, {\it Elements of Information
Theory},  2nd ed (Wiley, Hoboken, NJ) 2006.

\bibitem{ritort} E. H. Trepagnier, C.Jarzynski, F. Ritort, G. E. Crooks, C. Bustamante and J. Liphardt,  Proc. Nat. Acad. Sci. {\bf 101} 15038 (2005).


\bibitem{seki}
K. Sekimoto,  J. Phys. Soc. Jpn. {\bf 66}, 1234 (1997).

\bibitem{sekimoto}
K. Sekimoto,  Prog. Theor. Phys. Suppl. {\bf 130}, 17 (1998).


\bibitem{farago}
J. Farago,  J. Stat. Phys. {\bf 107}, 781 (2002); Physica A {\bf 331}, 69 (2004).

\bibitem{vanzon}
R. van Zon and E. G. D. Cohen, Phys. Rev. Lett. {\bf 91}, 110601 (2003).

\bibitem{blickle}
V. Blickle, T. Speck, L. Helden, U. Seifert and C. Bechinger,  Phys. Rev. Lett. {\bf 96}, 070603 (2006).

\bibitem{hfbt}
A. Gomez-Marin and J. M. Sancho,  Phys. Rev. E {\bf 73}, 045101(R) (2006).

\bibitem{visco}
P. Visco,  J. Stat. Mech.: Theory Exp. P06006 (2006).

\bibitem{maes2006}
M. Baiesi, T. Jacobs, C. Maes and N. S. Skantzos,  Phys. Rev. E  {\bf 74}, 021111 (2006).

\bibitem{joubaud}
F. Douarche, S. Joubaud, N.B. Garnier, A. Petrosyan and S. Ciliberto,  Phys. Rev. Lett.  {\bf 97}, 140603 (2006).

\bibitem{misawa} T. Misawa,  J. Math. Phys. {\bf 34}, 775 (1993).

\bibitem{misawa2} T. Misawa and H. Itakura, Phys. Rev. E {\bf 51}, 254 (1995).



\end{thebibliography}
\end{document}